\documentclass[fleqn,usenatbib]{mnras}

\usepackage{txfonts}

\usepackage[T1]{fontenc}
\usepackage{ae,aecompl}

\usepackage{graphicx}	

\usepackage{amsmath}	
\usepackage{amsfonts}   
\usepackage{amssymb}	
\usepackage{gensymb}    
\usepackage{soul} 
\usepackage{pdflscape}  

\usepackage{enumitem} 

\def \aj {AJ}
\def \mnras {MNRAS}
\def \pasp {PASP}
\def \apj {ApJ}
\def \apjs {ApJS}
\def \apjl {ApJL}
\def \aap {A\&A}
\def \nat {Nature}
\def \araa {ARAA}

\def \aaps {A\&A Suppl.}

\def \Msolar {\,M$_\odot$}

\newcommand{\hii} {H{\scriptsize{ II}} }

\def\lesssim{\mathrel{\hbox{\rlap{\hbox{\lower4pt\hbox{$\sim$}}}\hbox{$<$}}}}
\def\gtrsim{\mathrel{\hbox{\rlap{\hbox{\lower4pt\hbox{$\sim$}}}\hbox{$>$}}}}

\long\def\symbolfootnote[#1]#2{\begingroup%
\def\thefootnote{\fnsymbol{footnote}}\footnote[#1]{#2}\endgroup}

\title[Galactic Wolf-Rayet stars with \textit{Gaia} DR2 II]{Unlocking Galactic Wolf-Rayet stars with \textit{Gaia} DR2 II: Cluster and Association membership}

\author[Rate et al.]{
Gemma Rate,\thanks{garate1@sheffield.ac.uk} Paul A. Crowther, Richard J. Parker\thanks{Royal Society Dorothy Hodgkin Fellow}
\\
Department of Physics and Astronomy, University of Sheffield, Sheffield, S3 7RH, UK\\
}

\date{Accepted XXX. Received YYY; in original form ZZZ}

\pubyear{2019}

\begin{document}
\label{firstpage}
\pagerange{\pageref{firstpage}--\pageref{lastpage}}
\maketitle

\begin{abstract} 
 Galactic Wolf-Rayet (WR) star membership of star forming regions can be used to constrain the formation 
environments of massive stars. Here, we utilise {\it Gaia} DR2 parallaxes and proper motions to reconsider WR star membership 
of clusters and associations in the Galactic disk, supplemented by recent near-IR studies of young massive clusters. We find 
that only 18--36\% of 553 WR stars external to the Galactic Centre region are located in clusters, OB associations or obscured 
star-forming regions, such that at least 64\% of the known disk WR population are isolated, in contrast with only 13\% of O stars 
from the Galactic O star Catalogue. The fraction located in clusters, OB associations or star-forming regions rises to 25--41\% 
from a global census of 663 WR stars including the Galactic Centre region.  We use simulations to explore the formation processes 
of isolated WR stars. Neither runaways, nor low mass clusters, are numerous  enough to account for the low cluster membership 
fraction. Rapid cluster dissolution is excluded as mass segregation ensures WR stars remain in dense, well populated environments. Only low
density environments consistently produce WR stars that  appeared to be isolated during the WR phase. We therefore 
conclude that a significant fraction of WR progenitors originate in low density association-like surroundings which expand over 
time. We provide distance estimates to clusters and associations host to WR stars, and estimate cluster ages from isochrone fitting. 
\end{abstract}





\section{Introduction}
			

Classical Wolf-Rayet (WR) stars are the Helium core burning stage of the most massive O stars ($\geq$25\Msolar). Due to their short  lifetimes, they are excellent tracers of massive star formation and  evolution \citep{2007ARA&A..45..177C}.  In common with their O star progenitors, they have a strong influence on the galactic environment, ionizing \hii regions and 
expelling gas from their surroundings, which may both trigger and quench further star formation (e.g \citealt{2019ApJ...885...68B}).

Historically, the overwhelming majority of stars were thought to form in clusters \citep{2003ARA&A..41...57L}, which dissolve over time, although this is highly dependent on the definition of a 
cluster \citep{2010MNRAS.409L..54B}. Under the cluster formation 
scenario, progenitors of WR stars are believed to be restricted to relatively rich, dense clusters with masses in excess of $10^{3} M_{\odot}$ \citep{2010MNRAS.401..275W}, which are favourable to the formation of massive stars via competitive accretion or mergers. Therefore,  one would expect that WR stars are located in star clusters, unless they have been dynamically  ejected during the star formation process \citep{1967BOTT....4...86P} or following the core collapse supernova of a binary companion \citep{1961BAN....15..265B}.

However, it has recently been established that star formation in the nearby Cygnus OB2 association occurred at a relatively 
low density, with no evidence that massive stars formed in high density regions \citep{2014MNRAS.438..639W}. Indeed, from 
an inspection of v3 of the Galactic O star Catalogue \citep{2013msao.confE.198M}, only 42\% of O stars are thought to be members of star 
clusters, with almost three quarters located in OB associations and/or low density (<100s stars pc$^{-3}$) star forming regions (as summarised in Table~\ref{table:gosc}). 
These statistics are likely to be upper limits given membership has not been confirmed from \textit{Gaia} proper motions/parallaxes, although some comparisons with earlier distance estimates have been undertaken \citep{2019ApJ...882..180S}. 

\begin{table}
  \centering
  \caption[bf]{Summary of membership of clusters, OB associations of radio/infrared-selected star forming regions for
stars included in v3 of the Galactic O star Catalogue \citep{2013msao.confE.198M}.}
  \begin{tabular}{rrrrrr}
    \hline
Sample & Cluster &  OB Assoc & S.F. region & Isolated \\
       & Member  & Member & Member      & \\
    \hline
611  & 258 (42\%)        & 441 (72\%) & 445  (73\%)  &  82 (13\%) \\
  \hline
  \end{tabular}
  \label{table:gosc}
\end{table}


\newpage
For optically visible WR stars, \citet{1984A&AS...58..163L} found that only 10--30\% of stars identified at that time lay 
within clusters, which was updated by \citet{2001NewAR..45..135V} to include WR stars identified from infrared surveys, 
revealing that 35\% of Galactic WR stars were thought to be located either in clusters or associations. The known Galactic WR 
census has  grown substantially in recent years, with infrared surveys revealing significant populations in  clusters and the 
field  (\citealt{2011AJ....142...40M}; \citealt{2012AJ....143..149S}; \citealt{2013A&A...549A..98C}; 
\citealt{2015MNRAS.452.2858K}; \citealt{2018MNRAS.473.2853R}).

The second \textit{Gaia} data release, hereafter referred to as DR2, provides parallaxes, proper motions and positions for over a 
billion stars in the Galaxy \citep{2018AA...616A...1G}. Data from \textit{Gaia} supports the notion that not all stars are born in clusters 
\citep{2019arXiv191006974W} and increases the number of WR stars with trigonometric parallaxes from just one (WR11) to 383. In 
\citet{2020MNRAS.493.1512R} (Paper I), we used \textit{Gaia} parallaxes to calculate distances and new absolute magnitude 
calibrations for Galactic WR stars and to identify runaway candidates.

The environment of WR stars is also relevant to our understanding of core-collapse supernovae (ccSNe) which are deficient in hydrogen and/or helium; namely IIb, Ib and Ic, or collectively stripped-envelope supernovae (SE-SNe). Historically, massive WR stars were thought to be the immediate progenitors of SE-SNe. Over the last decade, evidence has accumulated suggesting the majority of SE-SNe arise primarily from lower mass stars in close binaries. This evidence includes ccSNe statistics \citep{2011MNRAS.412.1522S}, non-detection of WR progenitors in SE-SNe \citep{2013MNRAS.436..774E} and their inferred modest ejecta masses \citep{2011MNRAS.414.2985D, 2019MNRAS.485.1559P}.

In contrast, studies of ccSNe environments have established that those most stripped of hydrogen and helium are most closely associated with star-forming regions \citep{2012MNRAS.424.1372A, 2018A&A...613A..35K}. This is particularly acute for broad-lined Ic SNe and long Gamma Ray Bursts (GRBs) \citep{2006Natur.441..463F, 2012ApJ...759..107K} which are most closely linked to WR progenitors, as they possess the highest ejecta masses of SE-SNe \citep{2011ApJ...741...97D}; although long GRBs strongly favour low metallicities \citep{2008AJ....135.1136M}. Consequently, there is conflicting evidence for and against WR stars as the progenitors of highly stripped ccSNe, some of which directly involves their immediate environments.

Here we use the new distances, along with \textit{Gaia} proper motions, to analyse WR membership of Galactic clusters and 
associations, supplemented by IR surveys for sources inaccessible to \textit{Gaia}. We outline the methods in Section~\ref{sec:omem}. Cluster/association membership and distances are presented in Section~\ref{sec:mem}, ages are 
estimated in Section~\ref{ssec:agemem}. Finally, implications for massive star formation and their environments, informed by N-body simulations, are presented in Section~\ref{sec:massfm}. This is followed with a discussion and brief conclusions in Section~\ref{sec:con}.


\section{Assessment of cluster/association membership}\label{sec:omem}


\begin{table}
  \caption{Clusters and associations (in parentheses) excluded from \textit{Gaia} DR2 membership
analysis. Claimed membership from 
\citet{1984A&AS...58..163L}, 
\citet{2001NewAR..45..135V}, 
\citet{2012A&A...546A.110B}, 
\citet{2005AJ....130..126W}, 
\citet{2009ApJ...697..701M}, 
\citet{2015A&A...575A..10D}, 
\citet{2012MNRAS.419.1860D}, 
\citet{2008MNRAS.386L..23B}, 
\citet{2007A&A...475..209K} and 
\citet{2013A&A...549A..98C}.}
  \begin{tabular}{|p{.3\linewidth}|p{.3\linewidth}|p{.3\linewidth}|}
    \hline
    No reliable membership data & No parallaxes or proper motions from \textit{Gaia} (high $A_V$) & Few objects from membership list detected by \textit{Gaia} \\
    \hline
    AG Car         & Arches & C1104-610 a \\
    (Anon. Cen OB) & {[}DBS2003{]} 179 & C1104-610 b \\
    (Anon. Pup a)  & Galactic Centre & NGC 6871 \\
    (Anon. Pup b)  & Mercer 20 & (Serpens OB1) \\
    (Anon. Sct OB) & Mercer 70 &  (Serpens OB2) \\
    (Anon. Sco OB) & Mercer 81 & VVV CL099 \\
    (Anon. Vel a)  & SGR 1806-20 &  \\
    (Anon. Vel b)  &  Sher 1 & \\
    (Crux OB 4)    & Quartet & \\
    Dolidze 29    & Quintuplet & \\
    Henize 3       & VVV CL011 & \\ 
    (Norma OB4)     & VVV CL036 & \\
      (Vulpecula OB2)  & VVV CL073 & \\
                  & VVV CL074 & \\
                  & W43 & \\
                   & (Cas OB1) & \\
    \hline
  \end{tabular}
  \label{table:lost}
\end{table}

\subsection{Cluster/association candidates} \label{ssec:mlist}

The Galactic Wolf-Rayet catalogue \footnote{\url{http://pacrowther.staff.shef.ac.uk/WRcat/index.php}} includes 663 WR stars
(v1.23, July 2019) and lists the supposed members of star clusters and OB associations. To 
assess which WR stars are genuine
members  of a named cluster or association, we obtain lists of all candidate members from the literature and use these to determine 
the  proper motions and distances of the clusters and OB associations. The results were then compared to the proper motions and 
distances  of the individual WR stars. 

Of course, historical definitions of Galactic OB associations 
\citep{1978ApJS...38..309H} were undertaken from observations of visually bright O and B-type stars, so are inevitably
limited to stars located within a couple of kpc from the Sun. The majority of star clusters and OB associations are also
associated with optical nebulosities, drawn from one or more historical catalogues, namely the New General Catalogue \citep{1888MmRAS..49....1D}, Index Catalogue \citep{1910MmRAS..59..105D}, Sharpless 2 \citep{1959ApJS....4..257S} or RCW \citep{1960MNRAS.121..103R}.

Table \ref{table:gosc}, correlates stars in v3 of the Galactic O star Catalogue \citep{2013msao.confE.198M} to various regions to determine membership.  Numbers in different columns overlap because the O stars may be identified as members of both clusters and associations (due to hierarchical star formation) and an individual star may also have surrounding nebulosity in addition to cluster or association membership.

In contrast, only $\sim$7\% of the Galactic WR population detected by 
\textit{Gaia} lies within 2 kpc (Paper I), such that only a small fraction may lie within catalogued OB associations. Ideally,
membership of star-forming regions identified from radio \citep{2003A&A...397..133R} or infrared 
\citep{2004MNRAS.355..899C, 2010ApJ...719.1104R, 2014MNRAS.437.1791U} surveys would be more revealing, although this is beyond the 
capabilities of \textit{Gaia}. 

Consequently, here we focus on O and B star members of clusters and/or associations selected, where possible, to ensure 
a bright sample that could be reliably detected by \textit{Gaia} DR2 and fit the same distribution as our prior for WR stars 
(Paper I), (which results in somewhat lower distances  with respect to \citealt{2018AJ....156...58B}). Our prior consisted of a \hii region model, based on radio observations from \citealt{2004MNRAS.347..237P} and \citealt{2003A&A...397..213P}. This was combined with a dust disk model from \citet{2015MNRAS.447.2322R}. The dust was converted to an I band extinction map by calibrating the total dust along line of sight, with the maximum extinction at the Galactic centre. This map could be applied to the \hii region model, to approximate the \hii region distribution as observed by Gaia's white light G band. 

Unfortunately, some cluster  members lacked spectral type information. In these instances, we used 
the {\scriptsize{SIMBAD}} database to obtain the most 
recently assigned spectral type. However, many candidates remain unclassified. Additionally, for some larger candidate
catalogues, we used only the 20 brightest stars, as this provided a reasonable number of members for comparison and ensured these 
objects would be observed by \textit{Gaia}. Overall, we were able to use \textit{Gaia} data to assess the WR star membership in 28 clusters and 15 associations. We will revisit the issue of visually obscured WR stars in Section~\ref{sec:massfm}.

\subsection{Excluded clusters and associations} \label{ssec:excl}

Table ~\ref{table:lost} lists specific clusters and associations excluded from our \textit{Gaia} analysis. There are three 
main reasons why individual clusters and associations were omitted. No membership lists could be identified for Dolidze 29 or 
Henize 3. Anonymous associations in Cen, Pup, Sct, Sco and Vel, plus Norma OB4 and Crux OB4 also lacked membership information. 
Star lists were available for the parent region of Vulpecula OB2, but these did not break down into lists for specific OB 
associations.

The membership lists of other excluded clusters and associations were too small to test the WR membership, or were not  
available to \textit{Gaia}. Only 3 members of NGC 6871 were available in the \textit{Gaia} DR2 catalogue, including WR113, 
and the only stars detected by \textit{Gaia} for Sagittarius OB7, Serpens OB1 and Serpens OB2 were their supposed WR members.

The remaining clusters were not observed by \textit{Gaia}, as they are only accessible to IR observations, owing to high 
dust extinctions. For completeness, we include WR membership of embedded clusters in the Galactic disk, but only summarise previous 
results for the 110 WR stars within the Galactic Centre region ($l = 360\pm1^{\circ}$, $b = 0\pm 1^{\circ}$), which includes the 
Galactic Centre, Quintuplet and Arches clusters.


\begin{table*}
  \centering
  \caption{WR star membership of clusters for \textit{Gaia} DR2 sources (bold) and non 
\textit{Gaia} sources (non bold), external to the Galactic Centre region. (a) Decision was 
made based on proper motion and parallax 
clustering, not distances. (b) Large scatter in the data points. (c) Decision was made based on very few data points. (d) Possible former member ejected from cluster.}
  \begin{tabular}{|p{.12\linewidth}|p{.15\linewidth}|p{.15\linewidth}|p{.1\linewidth}|p{.08\linewidth}|p{.25\linewidth}|}
    \hline
    Cluster & Member & Possible member & Non member & References & Notes \\
    \hline
    Berkeley 86 &  &  & {\bf 139} & 1 & \\
    Berkeley 87 & {\bf 142} &  &  & 2  &  \\
    Bochum 7 & {\bf 12} &  &  & 3 & \\
    Bochum 10 & {\bf 23} &  &  & 4 & WEBDA. No spectral types. \\
    Bochum 14 & {\bf 104} &  &  & 3 & No spectral types. \\
    Cl 1813-178 & {\bf 111-4} & {\bf 111-2}$^{d}$ &  & 5 &  \\
    Collinder 121 &  & {\bf 6} &  & 3 & 20 brightest objects in the J band. Probability of membership>80\%. \\
    Collinder 228 &  & {\bf 24}$^{b}$ &  & 6 & GOSC. \\
    Danks 1 & {\bf 48a, 48-7} & {\bf 48-8, 48-9} &  & 7 & \\
    & {\bf 48-10} & {\bf 48-4} & & & \\
    Danks 2 & {\bf 48-2}$^{cb}$ &  &  & 7 &\\
    {[}DBS2003{]} 179 &  84-1, 84-6, 84-7 & &  &  8 & \\
    Dolidze 3 &  & {\bf 137}$^{ab}$ &   & 3 & 20 brightest objects in the J band. Probability of membership>80\%. \\
    Dolidze 33 &  & {\bf 120}$^{ab}$ &  & 3 & 20 brightest objects in the J band. Probability of membership>80\%. \\
    Havlen-Moffat 1 &  & {\bf 87}$^{d}$ & {\bf 89} & 6 & GOSC. \\
    Hogg 15 &  & {\bf 47}$^{c}$ &  & 4 & No spectral type information. \\
    Markarian 50 & {\bf 157} &  &  & 2 & \\
    Mercer 23 & {\bf 125-3}$^{a}$       &  &  & 9 & \\
    Mercer 30 & {\bf 46-3, 46-4, 46-5, 46-6} &  &  & 10 & \\ 
    Mercer 70 & 70-12  &  &   & 11 \\
    Mercer 81 & 76-2, 76-3, 76-4, 76-5, 76-6, 76-7, 76-8, 76-9 &   &    &  12 & \\ 
    NGC 3603 & {\bf 43-2, 42-1}$^{a}$ &  &  & 13 & \\
    & 43A, 43B, 43C  &  &  &  & WR43A, WR43B and WR43C unresolved by \textit{Gaia}.\\
    NGC 6231 &  & {\bf 79}$^{b}$, {\bf 79a}$^{b}$ & {\bf 78}  & 6 & GOSC. \\
    Pismis 20 &  & {\bf 67}$^{c}$ & & 2 & \\ 
    Pismis 24 &  &  & {\bf 93} & 6 & GOSC. \\
    Quartet   & 118-1, 118-2, 118-3 &  &   & 14 & \\ 
    Ruprecht 44 & {\bf 10} &  &   & 2 & \\
    SGR 0806--20 & 111a, 111b, 111c, 111d &   &    & 15 & \\
    Trumpler 16 & {\bf 25}$^{a}$ &  &   & 6 & GOSC. \\
    Trumpler 27 & {\bf 95, 98} &  &   & 2 & \\
    VVV CL009 & {\bf 45-5} &  &   & 16 & \\
    VVV CL036 & 60-6       &  &   & 16 & \\
    VVV CL041 & {\bf 62-2} &  &   & 17 & Selected cluster members, with J<16 \\
    VVV CL073 & 75-25, 75-26 &   &   & 16 & \\
    VVV CL074 & 75-27, 75-28, 75-29, 75-32 &   &    & 16, 18 & \\
    VVV CL099 & 84-8, 84-9, 84-10 &   &    & 16 &  \\
    W43       & 121a  &    &    &  19 & \\
    Westerlund 1$^{a}$ & {\bf 77aa, 77a, b} & {\bf 77p}$^{a,d}$, {\bf 77r}$^{d}$ &   & 20 & 
\\ 
    & {\bf 77c, d, f, h, i, j} & &  & & \\
    & {\bf 77m, n, o, q, s, sa} & &  & & \\
    & {\bf 77sb, sc, sd} & &  & & \\
    & 77e, g, k, l &  &  &  & \\
    Westerlund 2 & {\bf 20a, 20b}$^{a}$ &  &  & 21 & Stars with spectra (in table 3). \\
    \hline
  \end{tabular}
  (1) \citet{1995ApJ...454..151M}, 
   (2) \citet{2001AJ....121.1050M} and references therein, 
   (3) \citet{2014A&A...564A..79D}, \citet{2014yCat..35640079D} 
   (4) \citet{1990AJ.....99.2019L}, 
   (5) \citet{2011ApJ...733...41M}, 
   (6) \citet{2013msao.confE.198M}, 
   (7) \citet{2012MNRAS.419.1871D}, 
   (8) \citet{2012A&A...546A.110B},
   (9) \citet{2010A&A...516A..35H}, 
   (10) \citet{2016A&A...589A..69D},
   (11) \citet{2015A&A...575A..10D}, 
   (12) \citet{2012MNRAS.419.1860D},  
   (13) \citet{2008AJ....135..878M} ,
   (14) \citet{2009ApJ...697..701M}, 
   (15) \citet{2008MNRAS.386L..23B},  
   (16) \citet{2013A&A...549A..98C}, 
   (17) \citet{2015A&A...584A..31C}, 
   (18) \citet{2019A&A...627A.170M},
   (19) \citet{1999AJ....117.1392B},
   (20) \citet{2005A&A...434..949C}, 
   (21) \citet{2013AJ....145..125V}, 
  \label{table:possclust}
\end{table*}

\begin{table*}
  \centering
  \caption{Possible WR star membership of OB associations for \textit{Gaia} DR2 sources (bold) and non-\textit{Gaia} sources (non-bold), external to the Galactic Centre. (a) 
Decision was made based on pmra/pmdec and parallax clustering, not distances. (b) Large scatter in the data  points. (c) Decision was made based on very few data points. (d) 
Possible former member ejected from cluster, (e) Result taken directly from \citet{2019MNRAS.484.5834C}.}
  \begin{tabular}{|p{.15\linewidth}|p{.15\linewidth}|p{.15\linewidth}|p{.1\linewidth}|p{.08\linewidth}|p{.22\linewidth}|}
    \hline
    Association & Member & Possible member & Non-member & References & Notes \\
    \hline
    Ara OB1 &  &  & {\bf 77}$^{b}$ & 1 &  GOSC. \\
    Carina OB1 (incl Tr 16, Coll 232) & {\bf 22, 24, 25} & {\bf 18}$^{d}$, {\bf 23}$^{d}$ &  & 1 &  GOSC. \\
    Cassiopeia OB1 &  & 2$^{e}$&  & 2 &  No \textit{Gaia} astrometry.  \\
    Cassiopeia OB7 & & & {\bf 1}$^{c}$ & 1 &  GOSC. \\
    Centaurus OB1 & {\bf 48}$^{b}$ & & & 1 &  GOSC. \\
    Cephus OB1 &  & {\bf 152}$^{abc}$, {\bf 153}$^{abc}$ & & 1 &  GOSC. \\
    &  & {\bf 154}$^{abc}$, {\bf 155}$^{abc}$ & & &  \\
    Circinus OB1 &  & {\bf 67}$^{ab}$ & {\bf 65, 66, 68}$^{ab}$ & 3 & 20 brightest objects in the J band. No spectral types. Probability of membership>80\%. \\
    Cygnus OB1 & {\bf 137}$^{b}$, {\bf 138}$^{ab}$ &  & {\bf 136, 139} & 1 & GOSC. \\
    & {\bf 141}$^{b}$ & & & & \\
    Cygnus OB2 & {\bf 144, 145} & {\bf 142a}$^{d}$  &  & 1 & GOSC. \\
    Cygnus OB3 & {\bf 135}$^{b}$ &  & {\bf 134} & 1 & GOSC.\\
    Cygnus OB9 & {\bf 142a}$^{cb}$ &   &  & 1 & GOSC. \\
    Dragonfish & {\bf 46-2, 46-3, 46-4, 46-5, 46-6, 46-8, 46-9, 46-16, 46-17} &  & {\bf 46-10} & 4 &  \\
    Puppis OB2 & {\bf 10} &  &  & 5, 6 & \\
    Scorpius OB1 & {\bf 79, 79a} & {\bf 78}$^{d}$ &   & 1 & GOSC.  \\
    Sagittarius OB1 &  &  & {\bf 108, 104} & 1 & GOSC.  \\
    & & &{\bf 105}$^{bc}$, {\bf 110}$^{bc}$, {\bf 111}$^{bc}$ & &  \\
    Gamma Velorum & 11 & & & 7, 8 & WR 11 from Hipparcos catalogue. No spectral types. \\
    \hline
  \end{tabular}
  (1) \citet{2013msao.confE.198M}, 
  (2) \citet{2019MNRAS.484.5834C}, 
  (3) \citet{2013A&A...558A..53K}, 
  (4) \citet{2011ApJ...743L..28R}, 
  (5) \citet{2017MNRAS.472.3887M}, 
  (6) \citet{1981AJ.....86..222T},
  (7) \citet{2007A&A...474..653V},
  (8) \citet{2014yCat..35630094J}
  \label{table:possassoc}
\end{table*}

\subsection{Selection criteria} \label{ssec:cpar}

To assess cluster and association membership, we identified groups of stars by eye in RA and DEC proper motion space. We then compared this to the WR star proper motions, to determine if the latter were part of the groups. The \textit{Gaia} proper motion zero point is far smaller than the proper motion measurements ($\sim$10$\mu$as yr$^{-1}$, compared to mas scale proper motions, \citealt{2018A&A...616A...2L}) and therefore no corrections needed to be applied. Additionally, the uncertainties tended to be small when compared with parallax and distance uncertainties.

We assign individual WR stars as members of clusters/associations, possible members or non-members, depending on the 
similarity of proper motions with respect to other members. The proximity required for membership depends on the proper motion 
dispersions of the cluster or association. 4 WR stars in clusters and 4 WR stars in associations showed possible evidence of 
ejection, in which the star is located near the cluster or association in proper motion space (travelling within one or two mas 
yr$^{-1}$ in most cases), but is clearly isolated from the main group. It is possible that other clusters and associations could 
contain ejected stars, but these are masked by the scatter in the proper motion data. By way of example, 
\citet{2018MNRAS.480.2109D} support WR20aa and WR20c as potential stellar ejections from Westerlund~2 approximately 0.5 Myr ago.

Distances were used as a secondary check, to remove foreground and background stars. Parallaxes were converted to distances using 
the same prior and bayesian method as Paper I. The prior was based on \hii regions and extinction, and so is applicable to other OB 
cluster members. If WR stars showed strong agreement in proper motion space but disagreed in distance, they were assigned either as 
members or possible members, depending on how distant they were from the main cluster or association.

We checked our membership assignment was reasonable using the Python scikit-learn \citep{scikit-learn} implementation of DBSCAN. However, compared to the manual classification, the automated method had a number of limitations. When defining clusters in proper motion and parallax space, it struggled with boundary stars and could not account for sparse or scattered data. Additionally, it was difficult for this algorithm to properly weight the more reliable feature (proper motion) and account for quality indicators such as astrometric excess noise. We therefore chose not to use this automated method and used our manual classification to make the final membership decision.

As part of our analysis, we have obtained \textit{Gaia} DR2 distances to the clusters/associations. Although we could not model the shape and distance of each cluster simultaneously, (as recommended by 
\citealt{2018A&A...616A...9L}) it was still possible to get an approximation using the distances of individual members. To do this, we averaged positive parallaxes for all 
supposed members with astrometric excess noise <1. The systematic parallax uncertainty of the cluster or association could then be found by adapting the correlated error calculation outlined in 
(\citealt{gaia_pres}, \citeyear{2018A&A...616A...2L})
\begin{equation} \label{eq:dust_ext}
  \sigma_{clust} = \sqrt{\frac{1}{n}\langle\sigma_{\omega}^2\rangle+\frac{n-1}{n}\langle\langle V_{\omega}(\theta_{i,j} \rangle\rangle}
\end{equation}
where $n$ is the number of stars used to calculate the uncertainty, $\sigma_{\omega}$ (described in Paper I) is the inflated uncertainty of each star's parallax, averaged for the cluster. The $\sqrt{\frac{1}{n}\langle\sigma_{\omega}^2\rangle}$ term accounts for the random error and variance of the systematic error, using the external calibration with data from Table 1 of \citet{2018A&A...616A..17A}. However, it does not account for the spatial covariance function, $V_{\omega}(\theta)$, which is required to calculate the systematic errors for the mean parallax of stars in a cluster \citep{gaia_pres}. The full systematic term, $\frac{n-1}{n}\langle\langle V_{\omega}(\theta_{i,j}) \rangle\rangle$ (where $\langle\langle V_{\omega}(\theta_{i,j})\rangle\rangle)$ is the average $V_{\omega}(\theta)$ of n(n-1)/2 non redundant pairs of stars (i and j) in the cluster) is therefore required. 

The initial binned $V_{\omega}(\theta)$ values from \citet{2018A&A...616A...2L} were not sufficiently high enough resolution to account for the small angular separations between the stars in our clusters. We therefore fit a polynomial (with 14 parameters, although the results were not sensitive to changes in the number of these parameters) to the $V_{\omega}(\theta)$ data, in a similar manner to the bottom panel of figure 14 of \citet{2018A&A...616A...2L}. We then apply our prior from Paper I to the average parallax and uncertainty, to obtain the distance and its uncertainty.

In many cases, foreground or background objects had been misidentified as members and were contaminating the mean parallax. We 
therefore apply parallax cuts to remove these from the averages. We do not apply any cuts to associations (aside from removing a foreground star from Puppis OB2), as they may comprise multiple subregions, with different distances.


\section{WR membership}\label{sec:mem}

Table ~\ref{table:possclust} summarises WR membership of star clusters in the Galactic disk, drawn from \textit{Gaia} DR2 
proper motions (bold) or literature results for embedded clusters (non bold). Table~\ref{table:possassoc} provides a summary of WR 
membership of OB associations drawn from \textit{Gaia} DR2, supplemented by results for \citet{2019MNRAS.484.5834C} for WR2 (Cas 
OB1). 

Table ~\ref{table:possclust} reveals that only 43 WR stars from 62 claimed cluster members were confirmed from our 
analysis. Only 11\% of WR stars with \textit{Gaia} DR2 distances are in clusters. For associations, only 23 WR stars from 
48 claimed members were confirmed, including WR11 in the $\gamma$ Vel group (see Table ~\ref{table:possassoc}). However, 
membership of associations proved to be more challenging than clusters owing to greater scatter in proper motions and distances. 

Combining cluster and association membership (including WR25 and the members of Mercer 30, which are both association and cluster members, and WR24, WR79, WR79a and WR137 which are possible cluster members but confirmed association members), this rises to17\% of the total WR sample with \textit{Gaia} DR2 distances. Additionally, in many cases, only a few cluster/association members were detected by \textit{Gaia} DR2. This leaves the full proper motion and distance range of the cluster or association uncertain, which would potentially exclude WR members.

Several physically small or sparsely populated clusters, like Pismis 24 and Berkeley 86, were thought to host WR stars but do not. In the former's case, this is a cluster with few members and WR93 (WC7+O) has a radically different proper motion. \citet{1984A&AS...58..163L} only regarded it as a possible member and we can confirm it is not. For Berkeley 86, \citet{1984A&AS...58..163L} consider WR139 (WN5o+O) as a probable cluster member, but note it sits outside the apparent cluster and has a lower colour excess. We find that WR139 differs from known members in its proper motion and distance. Therefore we do not consider it a member.

A number of other clusters and associations did not have any confirmed members. This is because their proper motions are highly 
scattered, possibly because they are unbound, or broken down into subgroups along the line of sight. This made it difficult to 
locate the main proper motion centre of the cluster. For instance, Cassiopeia OB7 included a couple of possible members at a 
similar distance to WR1 (WN4b), but with no coherent proper motions.

Additionally, the existence of some clusters and associations is questionable. Ara OB1 shows a large scatter in proper motions, 
which indicates there is no relation between the supposed members. The catalogue for \citet{2013A&A...558A..53K} also suggests it 
may not be a cluster. Collinder 121 also contains stars with a wide range of proper motions, though they are all at approximately 
the same distance. Other clusters and associations with no or few members detected by \textit{Gaia}, such Serpens OB1, may also be 
chance alignments.

The proper motions of proposed WR members of Cir OB1 (WR65, WR67) agreed with other members; although their distances were in 
tension. In these instances future improvements to distance accuracy from \textit{Gaia}, would help with membership identification.

Tables~\ref{table:possclust}--\ref{table:possassoc} also include literature results for embedded clusters within the Galactic disk,
 which are inaccessible to \textit{Gaia}. Results are summarised in Table~\ref{table:summary}, and reveal that only 18\% of 553
WR stars in the Galactic disk are confirmed members of clusters or OB associations. 

OB associations included in the WR catalogue are nearby and have low associated extinction. However, the majority of WR stars are at $\sim$ kpc distances from the Sun and so beyond the extent of these catalogued associations. Additionally, more distant, moderately obscured star forming regions are historically detected at IR wavelengths but not at optical wavelengths and so these are not included in the OB associations of the WR star catalogue.

To account for this and incorporate these more distant star forming regions we have compared the location of WR stars to radio-selected \hii\ 
regions from  \citet{2003A&A...397..133R} and infrared selected star forming regions from \citet{2004MNRAS.355..899C}, 
\citet{2010ApJ...719.1104R}, and \citet{2014MNRAS.437.1791U}. In particular, \citet{2014MNRAS.437.1791U} provide star-forming complexes
from the Red MSX Source (RMS) survey of massive star forming regions within the Galactic disk. Accounting for potential membership of 
obscured star forming regions, the fraction associated with star clusters, OB associations or obscured star formation in the Galactic disk 
could be as high as $\sim36$\%.

If we include the 110 WR stars within the Galactic Centre region, of which 13 are members of the Arches cluster 
\citep{2018A&A...617A..65C}, 19 are members of the Quintuplet cluster \citep{2018A&A...618A...2C} and 36 lie 
within the Central Cluster \citep{1995ApJ...447L..95K, 2005ApJ...624..742T, 2006ApJ...643.1011P, 
2010ApJ...721..395F}, 25\% of 663 WR stars are confirmed members of clusters or associations, rising to 
41\% if potential association with radio/infrared star forming regions are confirmed.

\begin{table}
  \centering
  \caption[bf]{Summary of membership of clusters, OB associations and radio/infrared-selected star-forming regions (including candidates from Tables \ref{table:possclust} and \ref{table:possassoc}) for the known Galactic WR population. Some stars were members of both clusters and associations (where the cluster is a sub-region of the association), but we include these objects in the cluster statistics, as the cluster is their primary formation environment.}
  \begin{tabular}{rrrrrr}
    \hline
 Region &   Cluster & Assoc & Candid & Isolated & Total\\
    \hline
Disk \textit{Gaia} & 43 & 18 & 65 & 253 & 379 \\ 
Disk non-\textit{Gaia} & 37 & 1 & 37 & 99 & 174 \\ 
Galactic Centre & 68 &  & 2 & 40 & 110 \\ 
Total & 148 & 19 & 104 & 392 & 663 \\ 
  \hline
  \end{tabular}
   Additionally, some stars included in the original \textit{Gaia} distance total (WR11 and the stars in NGC3603) are here not counted as part of the disk Gaia population.
  \label{table:summary}
\end{table}

Tables ~\ref{table:otclust} and ~\ref{table:otassoc} compare our cluster and association distances with literature 
results. The distances of most clusters and associations are similar to previous estimates. However, 
we find that distances to Mercer 23, Mercer 30, Dolidze 3, Dolidze 33 and the Dragonfish association are closer than previous estimates. In particular, the revised distance of 5.2 kpc to the Dragonfish association is significantly closer 
than previous determinations of 12 kpc \citep{2016A&A...589A..69D} or 7 kpc \citep{2007A&A...475..209K}. However, the member stars 
are flagged with high (>0.3) astrometric excess noise and error to parallax ratios. This indicates the distance may dominated by 
the prior and therefore may be inaccurate. This is also relevant to its host cluster Mercer 30, which has a revised distance of 4.7 kpc, on the basis of just two members with positive parallaxes and astrometric excess noise below 1. 

Bochum 14 is found to be significantly more distant than previously thought.This distance is likely to be robust, 
since only one member has astrometric excess noise >0.3 (a further two members were removed for having astrometric excess noises 
above 1 mas).

We now discuss selected rich clusters/associations hosting multiple WR populations.

\renewcommand{\arraystretch}{1.5}
\begin{table*}
  \begin{center}
  \caption{Revised distances to star clusters using OB members obtained from \textit{Gaia} DR2 compared to literature values (indicated with DR2 if also obtained from \textit{Gaia}).}
  \begin{tabular}{|p{.15\linewidth}|p{.12\linewidth}|p{.09\linewidth}|p{.3\linewidth}|p{.08\linewidth}|p{.1\linewidth}|}
    \hline
    Cluster & Distance (this work) (kpc) & Number of stars & Previous distances (kpc) & References & Parallax cut (mas)\\
    \hline
    Berkeley 86 & 1.76$\substack{+0.09 \\ -0.08}$ & 11 & 1.91 & 1 & \\
    Berkley 87 & 1.72$\substack{+0.13 \\ -0.11}$ & 18 & 1.58 & 2 & \\
    Bochum 10 & 2.58$\substack{+0.24 \\ -0.20}$ & 8 & 2.7 & 3 & $\omega$<0.5 \\
    Bochum 14 & 2.88$\substack{+0.36 \\ -0.29}$ & 14 & 0.57 & 4 & $\omega$<0.5 \\
    Bochum 7 & 5.55$\substack{+1.02 \\ -0.78}$  & 21 & 5.6$\pm$1.7, 4.2$\pm$2.1 & 5, 6 & $\omega$<0.3 \\
    Cl 1813-178 & 2.05$\substack{+0.19 \\ -0.16}$ & 16 & 2.9$\pm$0.8 - 4.8$\substack{+0.25 \\ -0.28}$ & 7 &  \\
    Collinder 121 & 2.52$\substack{+0.14 \\ -0.13}$ & 9 & 0.75-1.00, 0.55 & 8, 9 & $\omega$<0.5 \\
    Collinder 228 & 2.54$\substack{+0.23 \\ -0.20}$ & 14 & 3.16, 2.01, 2.87 (DR2) & 2, 9, 10 &  \\
    Danks 1 & 3.41$\substack{+0.53 \\ -0.41}$ & 12 & 3.8$\pm$0.6 & 11 &  \\
    Danks 2 & 4.30$\substack{+0.73 \\ -0.57}$ & 5 & 3.8$\pm$0.6 & 11 & $\omega$<0.5 \\
    Dolidze 3 & 2.13$\substack{+0.17 \\ -0.15}$ & 21 & 1.03 & 4 &  \\
    Dolidze 33 & 2.96$\substack{+0.36 \\ -0.30}$ & 12 & 1.07 & 4 & $\omega$<0.5 \\
    Havlen-Moffat 1 & 3.13$\substack{+0.53 \\ -0.40}$  & 7 & 3.30 & 12 &  \\
    Hogg 15 & 3.20$\substack{+0.44 \\ -0.35}$ & 3 & 3.20 & 4 & $\omega$<0.3\\
    Markarian 50 & 2.52$\substack{+0.29 \\ -0.24}$  & 8 & 3.63, 3.46$\pm$0.35 & 2, 13 &   \\
    Mercer 23 & 3.36$\substack{+0.50 \\ -0.39}$ & 6 & 6.5$\pm$0.3 & 14 &  \\
    Mercer 30 & 4.72$\substack{+0.71 \\ -0.57}$ & 2 & 7.2$\pm$0.9, 12.6$\pm$1.5 & 15, 16 &  \\
    NGC 3603 & 6.74$\substack{+1.34 \\ -1.07}$ & 8 & 7.2$\pm$0.1 & 17 (DR2) & $\omega$<0.1 \\
    NGC 6231 & 1.60$\substack{+0.11 \\ -0.09}$ & 12 & 1.24 & 4 &  \\
    Pismis 20 & 3.44$\substack{+0.54 \\ -0.42}$ & 5 & 3.47, 3.18 & 2, 9 &  \\
    Pismis 24 & 1.71$\substack{+0.12 \\ -0.11}$ & 6 & 2.51 & 2 &  \\
    Ruprecht 44 & 5.38$\substack{+1.08 \\ -0.81}$ & 16 & 4.79 & 2 &  \\
    Trumpler 16 & 2.31$\substack{+0.22 \\ -0.18}$ & 16 & 3.16, 2.10, 2.87 (DR2) & 2, 9, 10 & $\omega$>0.3 \\
    Trumpler 27 & 2.43$\substack{+0.25 \\ -0.21}$ & 33 & 2.88 & 2 & $\omega$<0.5 \\
    VVV CL009 & 5.62$\substack{+1.27 \\ -0.94}$ & 6 & 5$\pm$1 & 18 &  \\
    VVV CL041 & 3.56$\substack{+0.59 \\ -0.46}$ & 18 & 4.2$\pm$0.9 & 19 &  \\
    Westerlund 1 & 3.78$\substack{+0.56 \\ -0.46}$  & 22 & 2.6$\substack{+0.6 \\ -0.4}$ (DR2), 3.87$\substack{+0.95 \\ -0.64}$ (DR2)  & 20, 21 & $\omega$<0.5 \\
    Westerlund 2 & 4.11$\substack{+0.80 \\ -0.59}$ & 21 & 4.16$\pm$0.07$\pm$0.26  & 22 & $\omega$<0.5 \\
    \hline
  \end{tabular}
  \label{table:otclust}
\end{center}
(1) \citet{1995ApJ...454..151M}, (2) \citet{2001AJ....121.1050M}, (3) \citet{2001MNRAS.325.1591P}, (4) \citet{2002A&A...389..871D},  
\citet{2014yCat....102022D}, (5) \citet{2018A&A...616A..40C}, (6) \citet{2007A&A...467..137C}, (7) \citet{2011ApJ...733...41M}, (8) 
\citet{2007ApJ...667L.155K}, (9) \citet{2017MNRAS.472.3887M}, (10) \citet{2019ApJ...882..180S}, (11) \citet{2012MNRAS.419.1871D}, 
(12) \citet{2001A&A...371..908V}, (13) \citet{2004MNRAS.355..475B}, (14) \citet{2010A&A...516A..35H}, (15) \citet{2007A&A...475..209K}, (16) \citet{2016A&A...589A..69D}, (17) \citet{2019MNRAS.486.1034D}, (18) \citet{2013A&A...549A..98C},  (19) \citet{2015A&A...584A..31C}, (20) \citet{2020MNRAS.492.2497A}, (21) \citet{2019MNRAS.486L..10D}, (22) \citet{2013AJ....145..125V}   
\end{table*}

\renewcommand{\arraystretch}{1.5}
\begin{table*}
  \begin{center}
    \caption{Revised distances to OB associations using OB members obtained
        from \textit{Gaia} DR2, compared to literature values (indicated with DR2 if also obtained from \textit{Gaia}).}
  \begin{tabular}{|p{.15\linewidth}|p{.12\linewidth}|p{.09\linewidth}|p{.3\linewidth}|p{.08\linewidth}|p{.1\linewidth}|}
    \hline
    Associations & Distance (this work) (kpc) & Number of stars & Previous distances (kpc) & References \\
    \hline
    Ara OB1a, b & 1.64$\substack{+0.05 \\ -0.05}$ & 9 & 1.3, 1.1/2.78 & 1, 2 \\
    Carina OB1     & 2.68$\substack{+0.18 \\ -0.16}$ & 82 & 1.8-2.8, 2.01, 2.87$\pm$0.73 (DR2) & 2, 3, 4 \\
    Cassiopeia OB1 & 2.4                             &  8 &                              & 5 \\
    Cassiopeia OB7 & 3.61$\substack{+0.17 \\ -0.16}$ & 3 & 2.01 & 2 \\
    Centaurus OB1 & 2.48$\substack{+0.10 \\ -0.09}$ & 9 & 1.92 & 2 \\
    Cephus OB1 & 3.40$\substack{+0.22 \\ -0.20}$ & 10 & 2.78 & 2 \\
    Circinus OB1 & 1.13$\substack{+0.03 \\ -0.03}$ & 24 & 2.01, 1.78 & 2 \\
    Cygnus OB1 & 1.97$\substack{+0.06 \\ -0.06}$ & 13 & 1.46 & 2 \\
    Cygnus OB2 & 1.57$\substack{+0.08 \\ -0.07}$ & 34 & 1.46 & 2 \\
    Cygnus OB3 & 2.05$\substack{+0.08 \\ -0.07}$ & 8 & 1.83 & 2 \\
    Cygnus OB9 & 1.62$\substack{+0.04 \\ -0.04}$ & 9 & 0.96 & 2 \\
    Dragonfish & 5.24$\substack{+0.89 \\ -0.69}$  & 12 & 12.4$\pm$1.7, 7.2$\pm$0.9 & 6, 7 \\
    Gamma Vel &  0.379$\substack{+0.004 \\ -0.004}$ & 20 & 0.345$\substack{+0.001+0.0124 \\ -0.001-0.0115}$-0.383$\substack{+0.0025+0.0153 \\ -0.0025-0.0142}$ (DR2) & 8 \\
    Puppis OB2 & 5.56$\substack{+0.55 \\ -0.46}$  & 8 & 3.18 & 2 \\
    Scorpius OB1 & 1.65$\substack{+0.07 \\ -0.07}$ & 26 & 1.53 & 2 \\
    Sagittarius OB1 & 1.21$\substack{+0.03 \\ -0.03}$ & 6 & 1.26 & 2 \\
    \hline
  \end{tabular}
  \label{table:otassoc}
\end{center}
(1) \citet{2011A&A...531A..73B}, (2) \citet{2017MNRAS.472.3887M}, (3) \citet{2016A&A...592A.149M}, 
(4) \citet{2019ApJ...882..180S}, (5) \citet{2019MNRAS.484.5834C}, (6) \citet{2016A&A...589A..69D}, (7) \citet{2007A&A...475..209K}, 
(8) \citet{2018A&A...616L..12F}
\end{table*}

\subsection{Carina nebula} \label{sssec:Carina}

The Carina nebula (NGC 3572) is the richest optically bright giant \hii\ region in the Milky Way. \textit{Gaia} DR2 confirms that Car OB1 hosts WR22, WR24 and WR25, with WR18 and WR23 possible members. The substructure of the region is quite complex (\citealt{2019A&A...622A.184B}, \citealt{2019MNRAS.486.4354R}), as 
it also contains the clusters Trumpler 16 and Trumpler 14 \citep{2016A&A...592A.149M}. WR25 is a member of Trumpler 16, which has a 
parallax of 0.430$\pm$0.115mas, corresponding to a distance of 2.18$\substack{+0.74 \\ -0.46}$ kpc. \citet{2018RNAAS...2c.133D} 
proposes a slightly larger parallax of 0.383$\pm$0.017mas, which falls within our uncertainties and \citet{2006ApJ...644.1151S} 
gives a distance of 2.35$\pm$0.05 kpc to $\eta$ Carinae/Trumpler 14. WR24 is also a possible member of Collinder 228, although this 
is difficult to confirm, as the cluster contains stars exhibiting a wide range of proper motions.

\citet{2016A&A...592A.149M} investigate the complex structure of Carina, identifying a foreground population at 
1.4--2.3 kpc (corresponding to Trumpler 18), a second population distributed over 2.0--3.3 kpc, plus a background group.
\citet{2019ApJ...882..180S} obtain 2.87 $\pm$ 0.73 kpc\footnote{Calculated using inverted parallaxes.} for 29 O star members of Trumpler 14--16 and Collinder 228 based on 
\textit{Gaia} DR2 parallaxes. We also note the  bulk of objects in our sample are between 2 and 4 kpc. \citet{2016A&A...592A.149M} also 
quote colour excesses of 0.3-0.6 mag. For our WR star sample, WR22, WR24 have values in this range, with 
E(B-V)=0.50$\pm$0.21 
and E(B-V)=0.35$\pm$0.21, respectively (Paper I). WR25 has a higher E(B-V)=0.93$\pm$0.31, using an anomalous reddening 
law of $R\substack{WR \\ v}=6.2$, from \citet{1995A&A...293..427C}. WR23 has a comparatively low E(B-V)=0.18$\pm$0.29 (Paper I) which suggests it could be a foreground object. However, the parallax derived distance is consistent with the Carina region and the reddening measurement has a large uncertainty.

\subsection{Cygnus OB2} \label{sssec:CygOB2}

\begin{figure}
  \includegraphics[width=\linewidth]{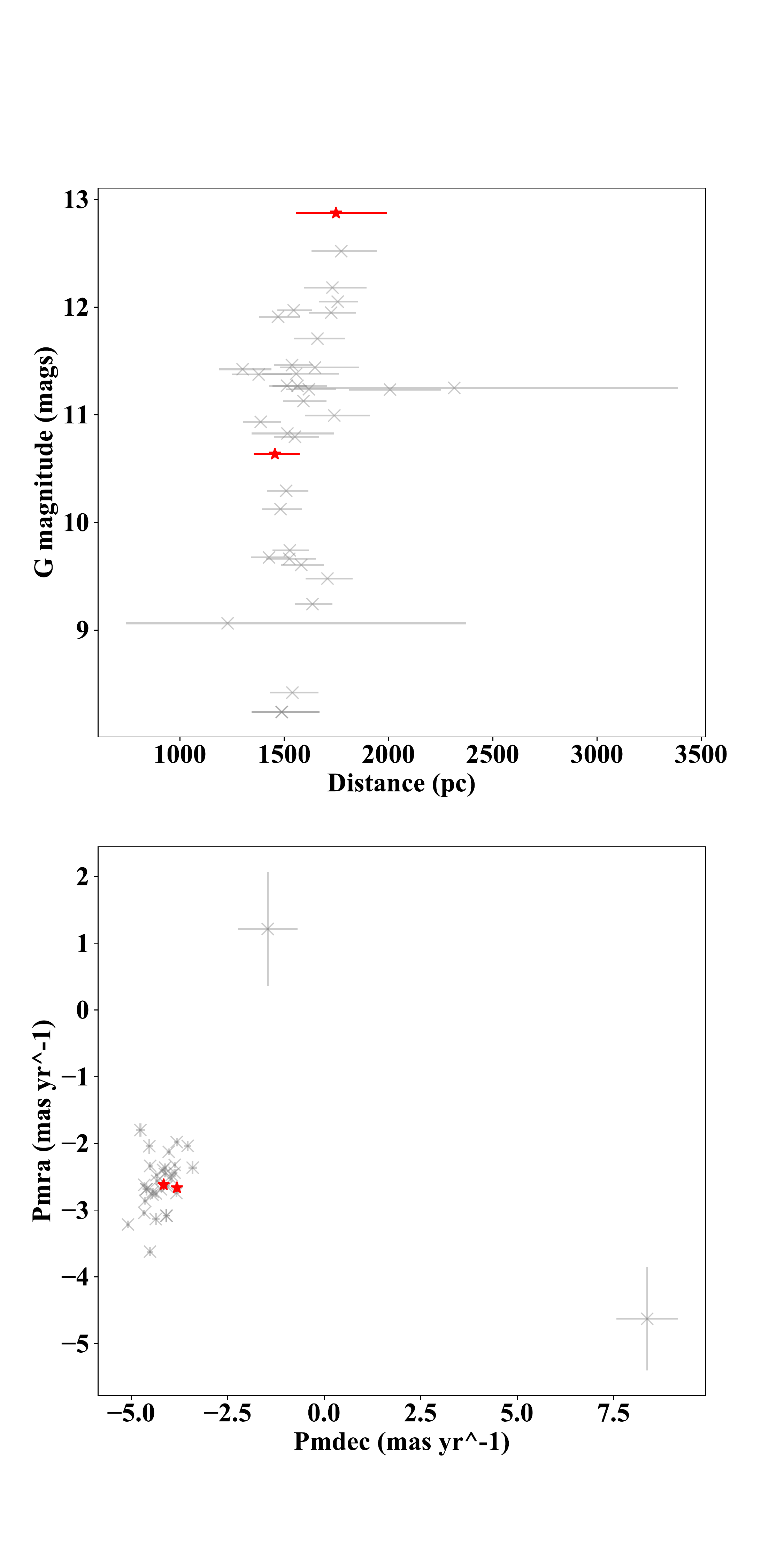}
  \caption{Distances vs G magnitudes (upper panel) and proper motions (lower panel) for members of Cyg OB2. Grey crosses are O and B stars from \citet{2013msao.confE.198M} while red stars are WR stars WR144 and WR145.} 
	\label{fig:cygob2}
\end{figure}

Cygnus OB2 is the nearest OB association rich in massive stars \citep{1991AJ....101.1408M}. We find a distance of 1.57$\substack{+0.08 \\ -0.07}$ kpc for Cygnus OB2, albeit with some substructure. Figure ~\ref{fig:cygob2} shows the distribution of distances and proper motions, indicating a spread from 1.4 kpc to 1.8 kpc (if uncertainties are included). WR144 is located towards the rear of the association at $\sim$1.7 kpc, whilst WR145 is closer to $\sim$1.4 kpc. Both these distances are in line with pre-\textit{Gaia} DR2 literature distances of 1.45 kpc \citep{2015MNRAS.449..741W} and 1.7 kpc \citep{1991AJ....101.1408M}.

\citet{2019MNRAS.484.1838B} have modelled the substructure of the cluster using DR2 data, and have  concluded that there are two main groups. One of these, at around 1.76 kpc, they term the 'main' group, with a 'foreground' group at 1.35 kpc. Our results place WR145 as a member of the foreground group and WR144 as a member of the more distant main group. 

\subsection{Danks 1 and 2} \label{sssec:Danks1}

Danks 1 and 2 clusters are young massive clusters within the G305 star formation complex \citep{2012MNRAS.419.1871D}. In Danks 1, three WR stars that were thought to be members have been confirmed, with three possible (but unconfirmed) members. Our membership list has very few entries for Danks 2, but we confirm WR48-2 is a member. The astrometric excess noise of all Danks 1 and 2 WR stars are greater than 0.3 mas, with WR48-4 exceeding 1 mas, indicating potentially unreliable astrometric results.

We find a distance of 3.41$\substack{+0.53 \\ -0.41}$ kpc  to Danks 1 and 4.30$\substack{+0.73 \\ -0.57}$ kpc to Danks 2, in fair agreement with the 3.8$\pm$0.6 kpc average distance of the G305 complex (hosting Danks 1 and 2), from \citet{2012MNRAS.419.1871D}. 

Danks 1 and 2 are in regions of high dust extinction, with $A_{K}$=1.1$\pm$0.16 for Danks 1 and $A_{K}$=0.92$\pm$0.29 for Danks 2 \citep{2012MNRAS.419.1871D}. This is consistent with $A_{K}$=0.99$\pm$0.22 for WR48-7 and $A_{K}$=0.83$\pm$0.20 for WR48-10 in Danks 1. However, WR48-2 in Danks 2 has $A_{K}$=0.48$\pm$0.20, significantly lower than the range for the cluster found by \citet{2012MNRAS.419.1871D}. In Paper I, we found the absolute magnitude for WR48-2 is anomalously faint for a WC7 or WC8 star, suggesting an underestimate of dust extinction, such that WR48-2 is a member of Danks 2.

\begin{figure}
  \includegraphics[width=\linewidth]{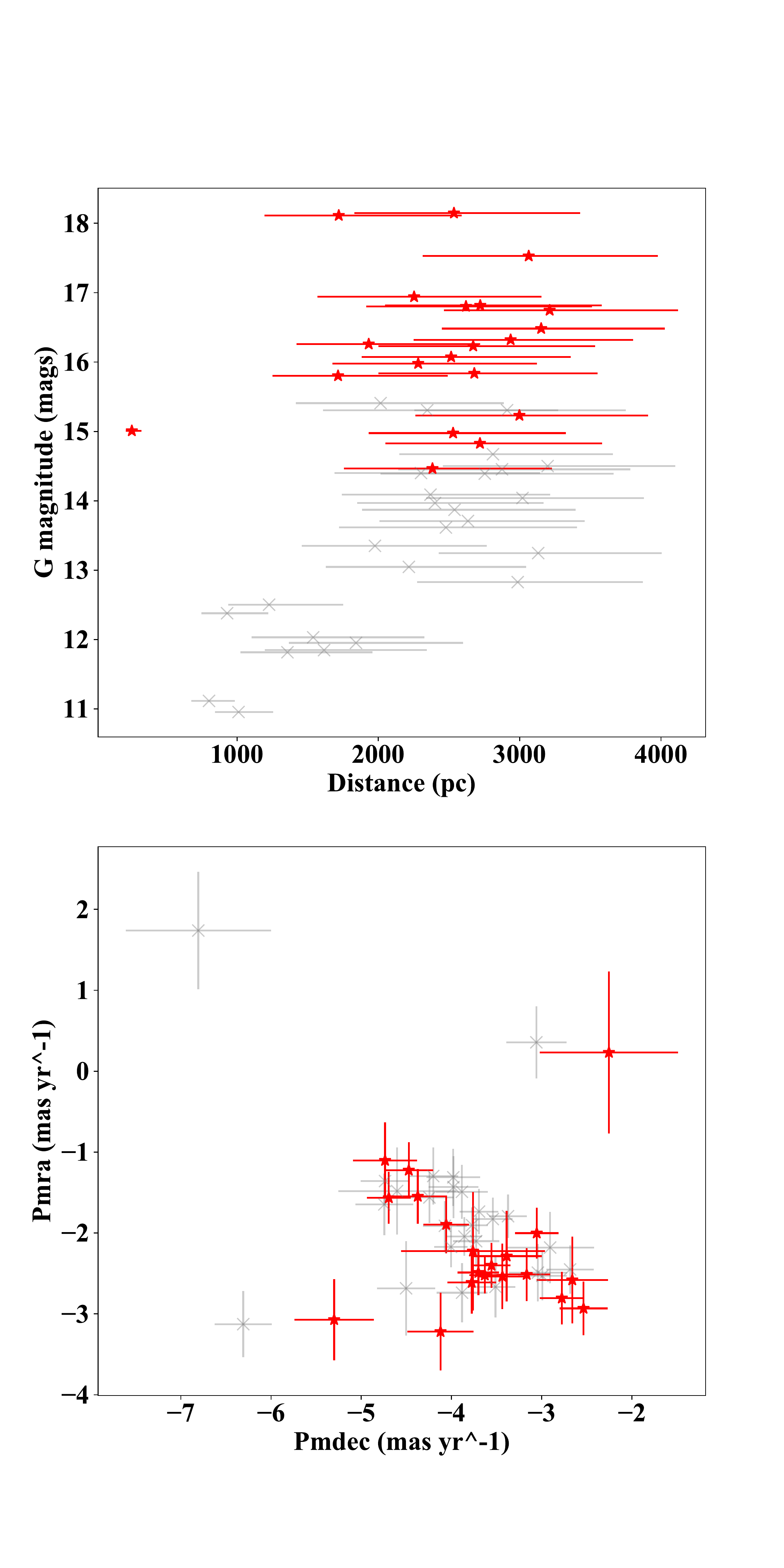}
  \caption{Distances vs G magnitudes (upper panel) and proper motions (lower panel) for members of Westerlund 1. Grey crosses are OB stars from \citet{2005A&A...434..949C} while red stars are WR stars from \citet{2006MNRAS.372.1407C}.} 
	\label{fig:Westerlund1}
\end{figure}

\subsection{$\gamma$ Velorum} \label{ssec:Gammavel}

WR11, the WC8 component of $\gamma$ Velorum, is confirmed as a member of its eponymous association. As WR11 is too bright for \textit{Gaia}, we use proper motion and parallax results from \textit{Hipparcos} \citep{2007A&A...474..653V} to confirm membership. The list of known members was compiled from the 20 brightest members in the V band \citep{2014yCat..35630094J}. These are not OB stars, because the $\gamma$ Velorum system is primarily surrounded by low mass stars \citep{2014A&A...563A..94J}.

The association has a wide range of proper motions, consistent with the suggestion by \citep{2014A&A...563A..94J} that it is barely bound. We find a distance of  0.379$\substack{+0.004 \\ -0.004}$ kpc to the group, consistent with the distance to WR11 (0.342$\substack{+0.038 \\ -0.030}$ kpc). It is also consistent with results from \citet{2018A&A...616L..12F}, who obtain two populations at 345.4$\substack{+1.0+12.4 \\ -1.0-11.5}$ pc and 383.4$\substack{+2.5+15.3 \\ -2.5-14.2}$ pc, respectively (accounting for both systematic and random errors). Two populations were also found by \citet{2014A&A...563A..94J}.

However, \citet{2019A&A...621A.115C} used \textit{Gaia} DR2 to find that the two populations previously identified (e.g \citealt{2014A&A...563A..94J}, \citealt{2018A&A...616L..12F}) are part of a much larger and more complex region around Vela OB2. These were possibly created by a supernova, which triggered star formation in the surrounding gas cloud.

Unfortunately, the current scope and methods of this paper do not allow us to fully disentangle the structure of Vela OB2 and assign membership to a specific subregion. We simply note that it is a member of this region and does therefore not appear to be isolated.

\subsection{NGC 3603} \label{sssec:NGC3603}

NGC 3603 is one of the youngest, most compact and high mass star clusters in the Milky Way \citep{2002ApJ...573..191M}. Unfortunately, young massive WN stars within the cluster core \citep{1998MNRAS.296..622C} could not be resolved with \textit{Gaia} DR2, but we confirm that WR42-1 and WR43-2 in its periphery are members.

We obtain a distance of 6.74$\substack{+1.34 \\ -1.07}$ kpc for NGC 3603, within uncertainty of the
literature values of 7.2 kpc \citep{1989A&A...213...89M} and 7.6 kpc \citep{2008AJ....135..878M}. Recently, \citet{2019MNRAS.486.1034D} obtained a distance of 7.2$\pm$0.1 kpc for a full sample of O star members, increasing to 8.2$\pm$0.4 kpc, when restricting the sample to stars within 1 arcmin of the cluster centre.

\subsection{Westerlund 1} \label{sssec:Westerlund1}

Westerlund 1 is an exceptionally rich star cluster \citep{2005A&A...434..949C}, thought to host 24 WR stars \citep{2006MNRAS.372.1407C}. \textit{Gaia} DR2 detects 20 of these stars, and we confirm 18 stars as members of Westerlund 1, as shown in Fig.~\ref{fig:Westerlund1}. In two cases membership could not be confirmed, owing to an unphysical distance (WR77p) or discrepant proper motions (WR77r). Unfortunately, many confirmed WR members  also have astrometric excess noises above 1 mas, which means their distances are somewhat unreliable. However, our primary membership indicator is proper motion, which is less vulnerable to large fractional uncertainties than parallax. A further 4 stars are not detected by \textit{Gaia}, which we assume to be members.

We estimate a cluster distance of 3.78$\substack{+0.56 \\ -0.46}$ kpc (though this excludes many stars with high excess noise). This is more distant than the 2.6$\substack{+0.6 \\ -0.4}$ kpc obtained from the full bayesian combination of cluster member parallaxes from \citet{2020MNRAS.492.2497A}.  The difference may stem from the fact we excluded some stars, via a parallax cut, as they seemed to be foreground objects. However, our result is consistent with 3.87$\substack{+0.95 \\ -0.64}$ kpc from \citet{2019MNRAS.486L..10D}. All three results from \textit{Gaia} are closer than the historical distance estimates of around 4--5 kpc 
\citep{2005A&A...434..949C, 2006MNRAS.372.1407C}. \citet{2019MNRAS.486L..10D} propose that the zero point is the dominant source of parallax uncertainty, adopting $-$0.05 mas instead of our $-$0.03 mas (Paper I).

\subsection{Westerlund 2} \label{sssec:Westerlund2}

Westerlund 2 is another rich, young high mass cluster \citep{2007A&A...463..981R}.
Proper motions for WR20a and WR20b  are comparable to the \textit{Gaia} cluster median of $\mu_{\alpha}=-5.172$ mas yr$^{-1}$, $\mu_{\delta}=2.990$ mas yr$^{-1}$ \citep{2018MNRAS.480.2109D}, favouring cluster membership. \citet{2018MNRAS.480.2109D} infer that WR20c and WR20aa possess proper motions consistent with recent (0.5 Myr) ejection from Westerlund 2. 

We obtain a distance to the cluster of 4.11$\substack{+0.80 \\ -0.59}$ kpc, which is close to the previous estimate of 4.16$\pm$0.07 (random)  +0.26 (systematic) kpc from \citet{2013AJ....145..125V}. There is some evidence for a background group or association \citep{2018MNRAS.480.2109D}, to which WR20a is a possible member (a distance of 5 kpc was inferred in Paper I). The extinctions of both WR stars are consistent with previous values for the cluster. \citet{2013AJ....145..125V} lists a range of 5.7<$A_V$<7.5 mag for OB stars in Westerlund 2, compared with $A_V$=6.44$\pm$0.64 and 7.57$\pm$0.64 mag for WR20a and WR20b, respectively, obtained from $A\substack{WR \\ v}$ (visual extinction in the Smith narrow band \citep{1968MNRAS.140..409S} in Paper I plus $A\substack{WR \\ v}$=$1.1A_V$ from \citealt{1982IAUS...99...57T}).


\section{Cluster ages} \label{ssec:agemem}


Armed with our revised distances and confirmed OB, WR members of star clusters, we are able to estimate ages from a comparison between cluster members and solar metallicity isochrones \citep{2011A&A...530A.115B}, following the approach of \citet{2001AJ....121.1050M} and \citet{2006MNRAS.372.1407C}. It is important to recognise that these isochrones are based on single stars and do not account for mass transfer in binaries, which may lead to resulting rejuvenated massive stars. Our results are therefore a lower limit for the true cluster ages (e.g \citealt{2020arXiv200402883S} find that single star isochrones can underestimate the true ages of \hii regions by 0.2 dex, when compared to binary population synthesis models).

Temperature calibrations for O stars are obtained from \citet{2005A&A...436.1049M}, whilst those for B stars are from \citet{2008flhs.book.....C}. O star bolometric corrections and intrinsic colours are from \citet{2006A&A...457..637M} (via \citealt{2006yCat..34570637M}). Intrinsic colours for B stars are taken from \citet{1994MNRAS.270..229W}. \citet{2006A&A...446..279C} provide bolometric corrections for supergiants in the V band, \citet{2007ApJS..169...83L} provide the same for dwarfs. 


The clusters Cl 1813-178, Danks 1 and Danks 2, Mercer 23, VVV CL009 and VVV CL041 were excluded from the age analysis, as only IR data was available for these clusters and spectral types for many O and B star cluster members were uncertain.

We categorise clusters with ages of $\leq$2 Myr as 'young', those with 2--5 Myr ages as intermediate and $\geq$5 Myr as old. Table~\ref{table:ages} lists cluster ages,  the adopted $R_V$ used to calculate reddening, average extinctions $A_V$ for cluster members, WR members and spectral types of OB stars within the cluster. Unfortunately, no spectral type information was available for the members of
Bochum 10 or 14 and so it was not possible to determine their cluster ages.

\begin{landscape}
  \begin{table} 
      \begin{center}
    \caption{Age estimates of star cluster within the Galactic disk host to WR stars, sorted by increasing age. Cluster membership
        of WR stars from \textit{Gaia} DR2 are indicated in bold. We categorise ages as either young ($\leq$ 2 Myr), intermediate (2--5 Myr) or old ($\geq$  5 Myr).}
    \begin{tabular}{|p{.1\linewidth}|p{.06\linewidth}|p{.06\linewidth}|p{.05\linewidth}|p{.05\linewidth}|p{.05\linewidth}|p{.1\linewidth}|p{.04\linewidth}|p{.15\linewidth}|p{.1\linewidth}|p{.06\linewidth}|}
      \hline 
      Cluster & Age in Myr (V-band) & Photo- metry Ref & $\mathrm{R_V}$ & $\mathrm{R_V}$ Ref & Mean $\mathrm{A_V}$ (mag) & Age in Myr (literature) & Age Ref & WR members (Sp Type) & OB Sp Type Range & OB Ref \\
      \hline 
      \multicolumn{11}{c}{--- Young --- } \\
      NGC 3603 & 1$\pm$1 & 2 & 3.55 & 1 & 4.9 & $1\pm1$, $1--4$  & 1, 2 & {\bf WR43-2} (O2If*/WN5), {\bf WR42-1} (WN4b), WR43A (WN6ha+WN6ha), WR43B (WN6ha), WR43C (O3If*/WN6) & O3V-O8.5V & 2 \\
      Trumpler 16 & 1$\pm$1 &  3, 4 & 3.1 & & 1.7 & 1.4 & 5 & {\bf WR25} (O2.5If*/WN6+O) & O3.5V-B0V & 3, 4\\
      Westerlund 2 & $2\pm1$ & 6 & 4.1 & 6 & 7.2 & $<1$ & 6 & {\bf WR20a} (O3If*/WN6+ O3If*/WN6), {\bf WR20b} (WN6ha) & O3V-O8V & 7, 8 \\
      Collinder 228 & $\sim2$ & 5 & 3.1 &   & 1.4 & &  & {\bf WR24} (WN6ha) & O5III-9.5V & 5  \\
      \multicolumn{11}{c}{--- Intermediate --- } \\
      Westerlund 1 & $<5$ & 9 & 3.1 & & 12.6 & 4.5-5 & 10 & {\bf WR77aa} (WC9d), {\bf WR77a} (WN6) &  O9III-B5Ia & 9 \\ 
      Bochum 7 & $\sim5$ & 11 & 3.3 & 12 & 2.7 & $<3$ & 11 & {\bf WR12} (WN8h) & O6.5V-B0V & 11  \\
      \multicolumn{11}{c}{--- Old --- } \\
      Ruprecht 44 & $7\pm3$ & 13, 5 & 3.1 & & 1.9 & 3 & 5 & {\bf WR10} (WN5ha) & O8III-B1V & 13, 5 \\
      Trumpler 27 & $7\substack{+3 \\ -2}$ &  14, 5 & 3.1 & 14 & 4.7 & 5 & 5 & {\bf WR95} (WC9d), {\bf WR98} (WN8/C7) & O8III-B8I & 15, 5 \\
      Berkeley 87 & 8--9 &  15, 5 &  3.1 & & 5.1 & 3 & 5 & {\bf WR142} (WO2) & O8.5III-B1V & 16, 5  \\
      Markarian 50 & $\sim10$ &  5 &  3.1 & &  2.5 & 7.4 & 5 &  {\bf WR157} (WN5-B1II) &  B0III-B1.5V & 17, 5 \\
      \hline
    \end{tabular}
    \label{table:ages}
    \end{center}
  (1) \citet{2004AJ....127.1014S}, (2) \citet{2008AJ....135..878M}, (3) \citet{1993AJ....105..980M}, (4) \citet{2006MNRAS.367..763S}, \citet{2006MNRAS.368.1983S}, (5) \citet{2001AJ....121.1050M}, (6) \citet{2015MNRAS.446.3797H}, (7) \citet{2007A&A...463..981R}, (8) \citet{2011A&A...535A..40R}, (9) \citet{2020A&A...635A.187C}, (10) \citet{2006MNRAS.372.1407C}, (11) \citet{2018A&A...616A..40C}, \citet{2018yCat..36160040C}, (12) \citet{1999JKAS...32..109S}, (13) \citet{1981AJ.....86..222T}, (14) \citet{2012A&A...548A.125P}, (15) \citet{1977ApJ...215..106M}, (16) \citet{1982PASP...94..789T}, (17) \citet{1983AJ.....88.1199T}
  \end{table}
  \end{landscape}

All four young clusters host hydrogen-rich main sequence WN stars, Of/WN stars and early O dwarfs. These add to the increasing evidence that main-sequence very massive stars exhibit transition Of/WN or weak-lined WNh spectral morphologies \citep{2010MNRAS.408..731C}. Two clusters have age estimates of $\leq$5 Myr, so one would expect them to host classical WR stars (hydrogen-deficient WN and WC stars) and mid-type O stars. This is true in both instances and although the age estimate for Westerlund 1 is an upper limit, it is in line with the previous literature value from \citet{2006MNRAS.372.1407C}.

Four clusters were assigned ages of $>$5 Myr, owing to the presence of late O and early B giants. Both Markarian 50 (WR157) and Ruprecht 44 (WR10) also host weak-lined WN5 stars, with previous age estimates for Markarian 50 pointing to ages of 7 Myr or greater \citep{2001AJ....121.1050M, 2006MNRAS.372.1407C}. In contrast, previous estimates of the age of Ruprecht 44 have indicated $\sim$3 Myr, and the luminosity of WR10 from \citet{2019A&A...625A..57H} adjusted to the revised distance of Paper I is relatively high ($\mathrm{log(L/L_{\odot})}$=5.8) for an age of 7 Myr, suggesting it could be a rejuvenated merger product \citep{2014ApJ...780..117S}.

Trumpler 27, another old cluster, contains a very late WC and WN/C type; both expected for a younger cluster. However, the uncertainty of the result may mean that the cluster is somewhat younger than 7 Myr, with the younger limit of 5 Myr more in line with previous estimates. The WO star in Berkeley 87 would also appear to be too young for a cluster 8--9 Myr old. However, we note that there is a wide scatter in the possible isochrones that would fit this cluster. The bulk of stars are best fit to the 8--9 Myr isochrone, but there are two outliers at $\sim$4 Myr (including the only O star in the sample). These younger outliers better match the age result from \citet{2001AJ....121.1050M} and may indicate the presence of multiple populations in Berkeley 87.

For the selected star clusters, there is no evidence of a population of low luminosity WR stars in old clusters, originating from envelope stripping by a close companion. Models from \citet{2018A&A...615A..78G} indicate that stripped helium stars can exhibit a variety of spectral types depending on their mass. At solar metallicity, progenitors of $\geq15 M_{\odot}$ mass donors in close binaries which produce stripped helium stars of $\geq 5 M_{\odot}$ are anticipated to resemble WN stars. However, due to their low luminosities and/or dilution from an early-type companion (mass gainer), these features may go undetected.


\section{Implications for massive star formation and environments}\label{sec:massfm}

We have confirmed literature results of a low cluster membership fraction of 14\% for WR stars within 
the Galactic disk, increasing to at most 36\% after OB association/star-forming region membership is considered 
(Table~\ref{table:summary}). If the O-type progenitors of WR stars primarily originated in populous, or high mass star clusters, the only way to produce such a low WR cluster membership fraction is if the WR stars are ejected from the cluster, or if the cluster dissolves and is consequently unrecognisable. Only 42\% of nearby Galactic O stars currently lie within known star clusters (Table~\ref{table:gosc}), so assuming a low ejection fraction or modest ejection velocities, approximately half of WR progenitors formed within clusters have been lost. A relatively high fraction of O stars can be dynamically ejected from dense, relatively  massive star clusters over the first few Myr \citep{1967BOTT....4...86P}, albeit with relatively modest velocities, of order 10 km\,s$^{-1}$ \citep{2016A&A...590A.107O}. 

Alternatively, the majority of WR progenitors may originate in OB associations, but be ejected following the disruption of their binary systems due to core-collapse supernovae \citep{1961BAN....15..265B}. Over 70\% of Galactic OB stars in the Solar 
Neighbourhood (few kpc) are found in OB associations/star forming regions, whereas at most 36\% of WR stars external to the 
Galactic Centre region are associated with a star forming region. Since the majority of massive stars appear to be born in close 
binary systems \citep{2012Sci...337..444S}, it is possible that WR stars are ejected through this mechanism. However, simulations 
suggest only 3\% of such binaries lead to runaway WR stars, with $\geq$30 km\,$^{-1}$ \citep{2013MNRAS.436..774E}, with 
slower moving walkaway stars much more common \citep{2019A&A...624A..66R}.

It is therefore apparent that WR stars may be ejected either dynamically from dense clusters, or via the disruption of a binary system following a supernova (albeit with relatively modest velocities in most instances). Recalling 1 km\,s$^{-1}$ equates to 1 pc/Myr, a WR star with an age of 4 Myr moving at 10 km\,s$^{-1}$ would travel no more than 40 pc from its birth site, usually much less owing to the 
delayed timescale for dynamical ejection/binary disruption. In contrast, field WR stars dominate the population in the Galactic disk, with runaways relatively common. From Paper I, we identified 8\% of WR stars from \textit{Gaia} DR2 to lie at least three \hii\ scale heights from the Galactic midplane, representing a minimum runaway fraction. The true runaway fraction must be higher, since these statistics neglect WR stars ejected within the disk. Indeed, the runaway fraction of O stars is 10--25\% \citep{1986ApJS...61..419G}, with a high fraction of runaways amongst the field O star population \citep{2005A&A...437..247D}.

The fact that only a minority of O stars are found in open clusters, together with the tension between the scarcity of 
predicted fast moving WR stars from low-mass clusters/close binaries, and the observed runaway fraction from the field population of WR stars, argues for an alternative to the usual assumption that their progenitors originate in dense clusters. In the following subsections we consider 
the possibility that WR progenitors originate in low density star-forming regions which are not recognised as clusters/associations, or that their host star cluster has dissolved.

We do not include primordial binaries in these simulations, despite the high incidence of close binaries of massive stars \citep{2012Sci...337..444S}. If binary systems containing massive stars form via capture, then we implement stellar and binary evolution. However, we do not find any instances of the formation of very close binaries that would subsequently undergo common-envelope evolution in our simulations (and hence lengthen the WR-phase).

The close binary channel may produce main sequence mergers or strip the envelope of the 
primary through Roche Lobe overflow \citep{2014ApJ...782....7D}, extending the limit for the formation of WR stars to lower masses. \citet{2020A&A...634A..79S} suggest a lower initial mass threshold of 18$M_{\odot}$ for solar-metallicity WR stars, while \citet{2018A&A...615A..78G} suggest 15 $M_{\odot}$, although such helium core masses/luminosities lie below those of WR stars included in our study. Nevertheless, they are considered to dominate the statistics of stripped envelope core-collapse supernovae (\citealt{2013MNRAS.436..774E}; 
\citealt{2011MNRAS.412.1522S}).

The simulations presented below thus assume that WR stars originate from (initially) single stars with masses in excess of 25 $M_{\odot}$. According to \citet{2005A&A...429..581M} the lower mass limit to the formation of single WR stars at solar metallicity is 22 $M_{\odot}$ for rapid rotators, or 37 $M_{\odot}$ for non-rotators, while \citet{2020A&A...634A..79S} obtain 20--30 $M_{\odot}$. 

\subsection{High mass stars in low-mass clusters and associations} \label{ssec:low_mass}

\begin{figure}
  \begin{center}
\includegraphics[scale=0.32]{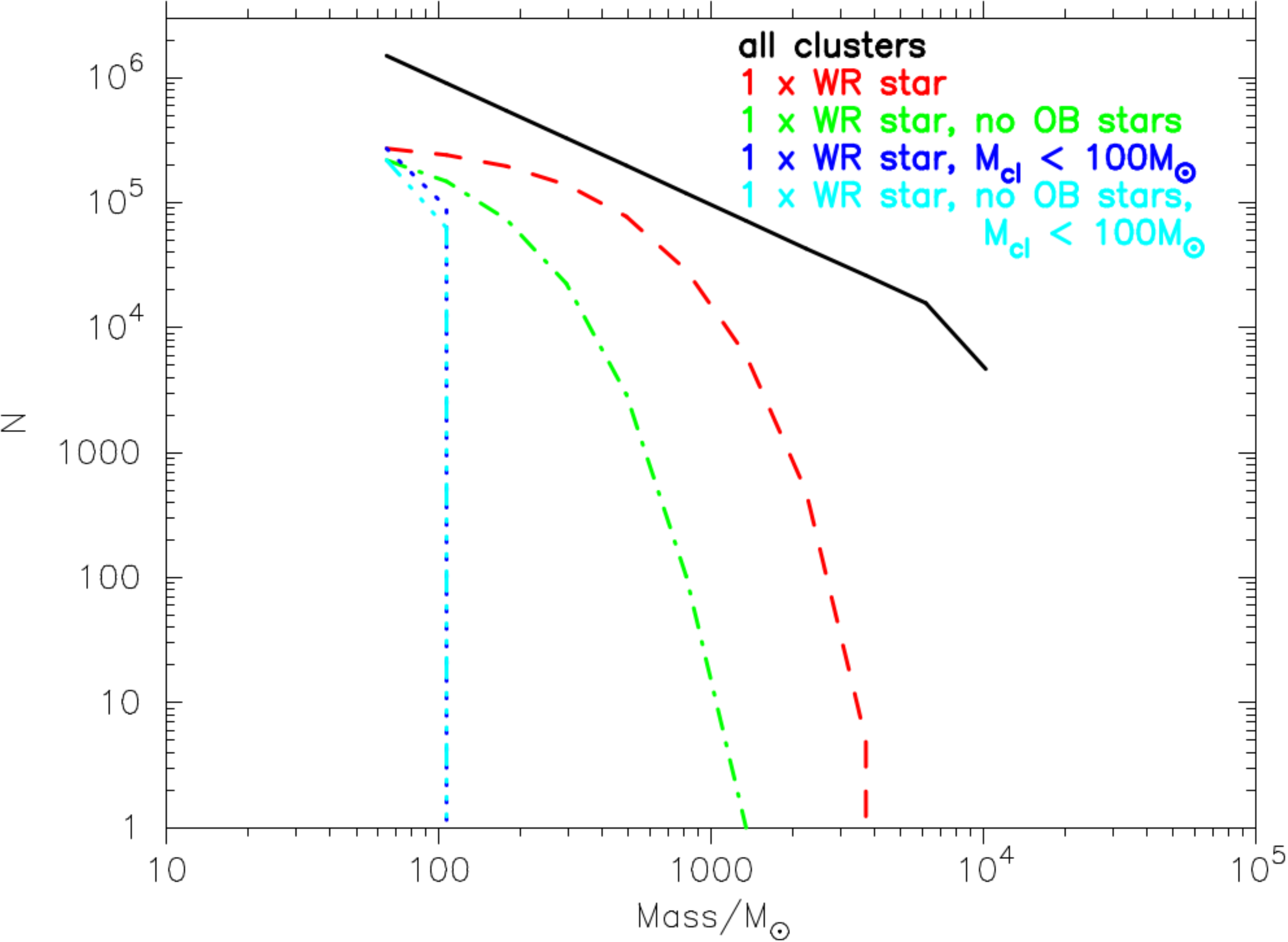}
\includegraphics[scale=0.32]{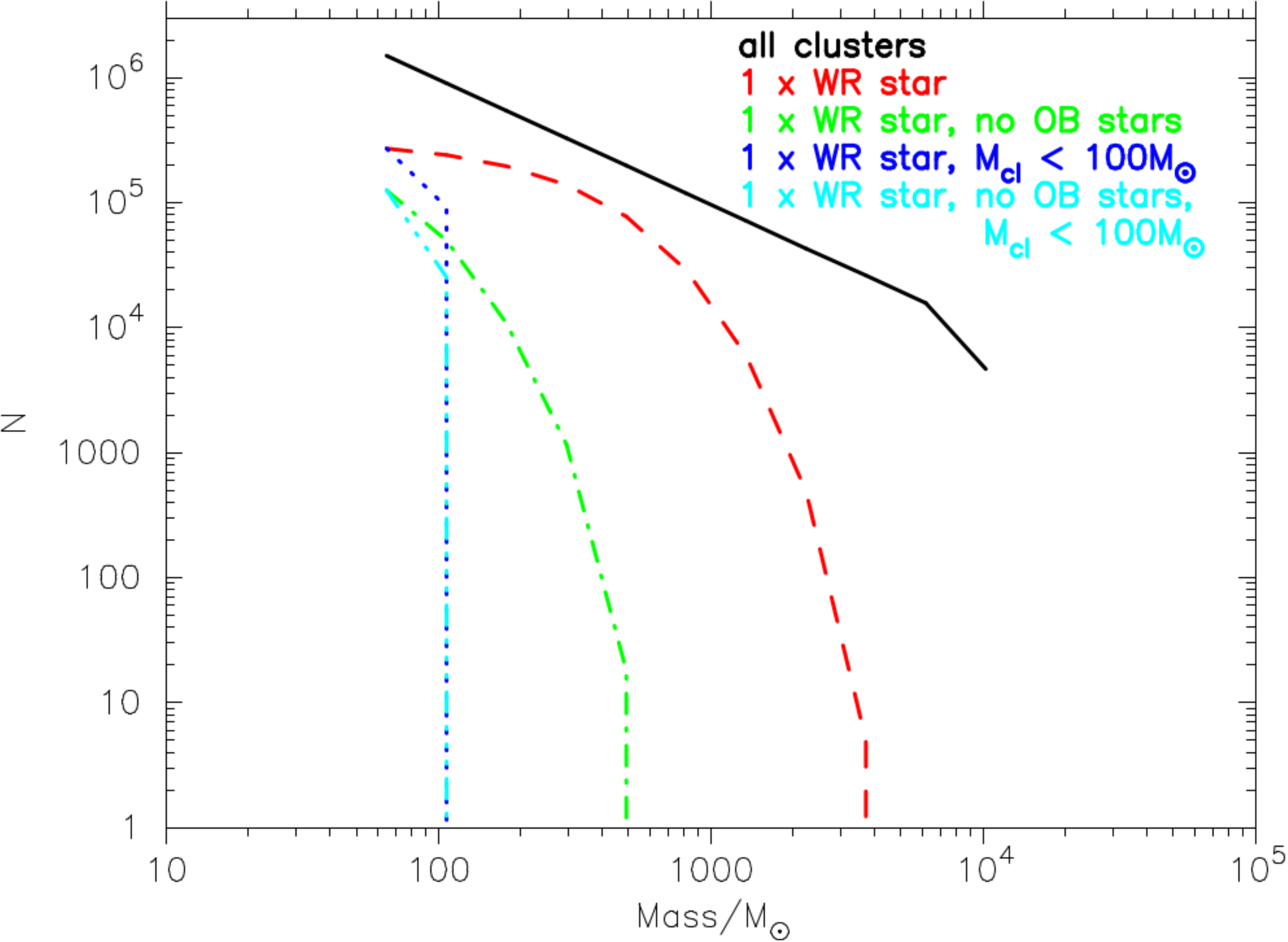}
\caption{Monte Carlo simulations of the number of clusters with different masses, which contain WR stars. Each line shows the relation between cluster mass and number if they contain one WR star. The cutoff applied in the upper panel was >10 \Msolar\ to OB stars (corresponding to O stars and early B stars, which are the brightest OB subtypes) and >25 \Msolar\ for WR stars. For the lower panel the cutoff was >5 \Msolar\ for OB stars (a stricter criterion removing O stars, early B and mid B stars) and >25 \Msolar\ for WR stars.}
\label{fig:Obcomp}
  \end{center}
\end{figure}

Here we consider the possibility that an apparently isolated WR star is in fact part of a low-mass star-forming region 
that has formed one massive star, with the remaining stellar content too faint to be observed \citep{2007MNRAS.380.1271P}.

One piece of evidence for WR stars in such environments comes from isolated protostellar cores of high mass stars. An example is 
G328.255--0.532 which may eventually form a $\sim$50\Msolar O star \citep{2018A&A...617A..89C}.

A notable example of an existing low-mass star forming region host to massive stars is the $\gamma$ Velorum group 
\citep{2014A&A...563A..94J},  Additionally, \citet{2016A&A...589A..70P} finds that the total mass of the cluster is only $\sim$100 
\Msolar. However, accounting for the wider environment, \citet{2019A&A...621A.115C} find a total mass of 2330$M_{\odot}$. This potentially indicates that these regions surrounding a single massive star appear low mass, but may be part of much wider, more massive star forming regions.

In order to test the hypothesis that our observed isolated WR stars are the most massive stars within low-mass, faint regions, we perform a Monte Carlo experiment similar to those in \citet{2007MNRAS.380.1271P} and \citet{2019MNRAS.484.2692T}. First, we sample cluster masses, $M_{\rm cl}$ in the range $50 - 10^4$M$_\odot$, from a single power-law of the form
\begin{equation}
N(M_{\rm cl}) \propto M_{\rm cl}^{-\beta},
\end{equation}
where $\beta = 2$ \citep{2003ARA&A..41...57L}. Once the cluster mass has been selected, we  populate the cluster with stellar masses drawn from a \citet{2013MNRAS.429.1725M} Initial Mass Function, which has a 
probability density function of the form
\begin{equation}
p(m) \propto \left(\frac{m}{\mu}\right)^{-\alpha}\left(1 + \left(\frac{m}{\mu}\right)^{1 - \alpha}\right)^{-\beta},
\label{maschberger_imf}
\end{equation}
where $\mu = 0.2$\,M$_\odot$ is the average stellar mass, $\alpha = 2.3$ is the \citet{1955ApJ...121..161S} power-law exponent for higher mass stars, and $\beta = 1.4$ 
describes the slope of the IMF for low-mass objects \citep*[which also deviates from the log-normal form;][]{2010ARA&A..48..339B}. We sample this distribution in the mass 
range 0.1 -- 300\,M$_\odot$, which allows for the most massive stars known to form \citep{2010MNRAS.408..731C}.

We sample from these distributions until we obtain a total stellar mass of $10^9$M$_\odot$. We then determine how many clusters 
contain one WR star (defined as having an individual mass $>$25\,M$_\odot$, by assuming solar metallicity 
\citep{2007ARA&A..45..177C} and no other OB stars. These are defined as having individual masses $>$5\,M$_\odot$ (to exclude O 
stars, early and mid B-type stars, where the latter are the most faint OB spectral types likely to be visible) or $>$10\,M$_\odot$ 
(to exclude O stars and early B stars, which would be visible in most cases).

Figure~\ref{fig:Obcomp} shows the cluster mass functions for all clusters (solid black line), clusters containing exactly one WR 
star (red dashed line), clusters containing one WR star and no other OB stars (green dot-dashed line), clusters containing one WR 
star with the remaining stellar mass $<$100\,M$_\odot$ (the dark blue dotted line) and clusters containing one WR star, no OB stars 
and with the remaining stellar mass $<$100\,M$_\odot$ (the cyan dot-dashed line).

We consider that only WR stars within clusters for which the remaining stellar mass $<$100\,M$_\odot$ could be 
mis-classified as being isolated. The upper and lower panels of Figure~\ref{fig:Obcomp} show that these low mass (<100 \Msolar) 
clusters containing 1 WR star and no OB stars (defining OB stars as >10 \Msolar\ and >5 \Msolar, respectively), will only form in 
around 8\%--15\% of instances. This is significantly smaller than our observed isolated fraction of 59--75\% of WR stars.

\begin{figure}
  \begin{center}
\includegraphics[scale=0.38]{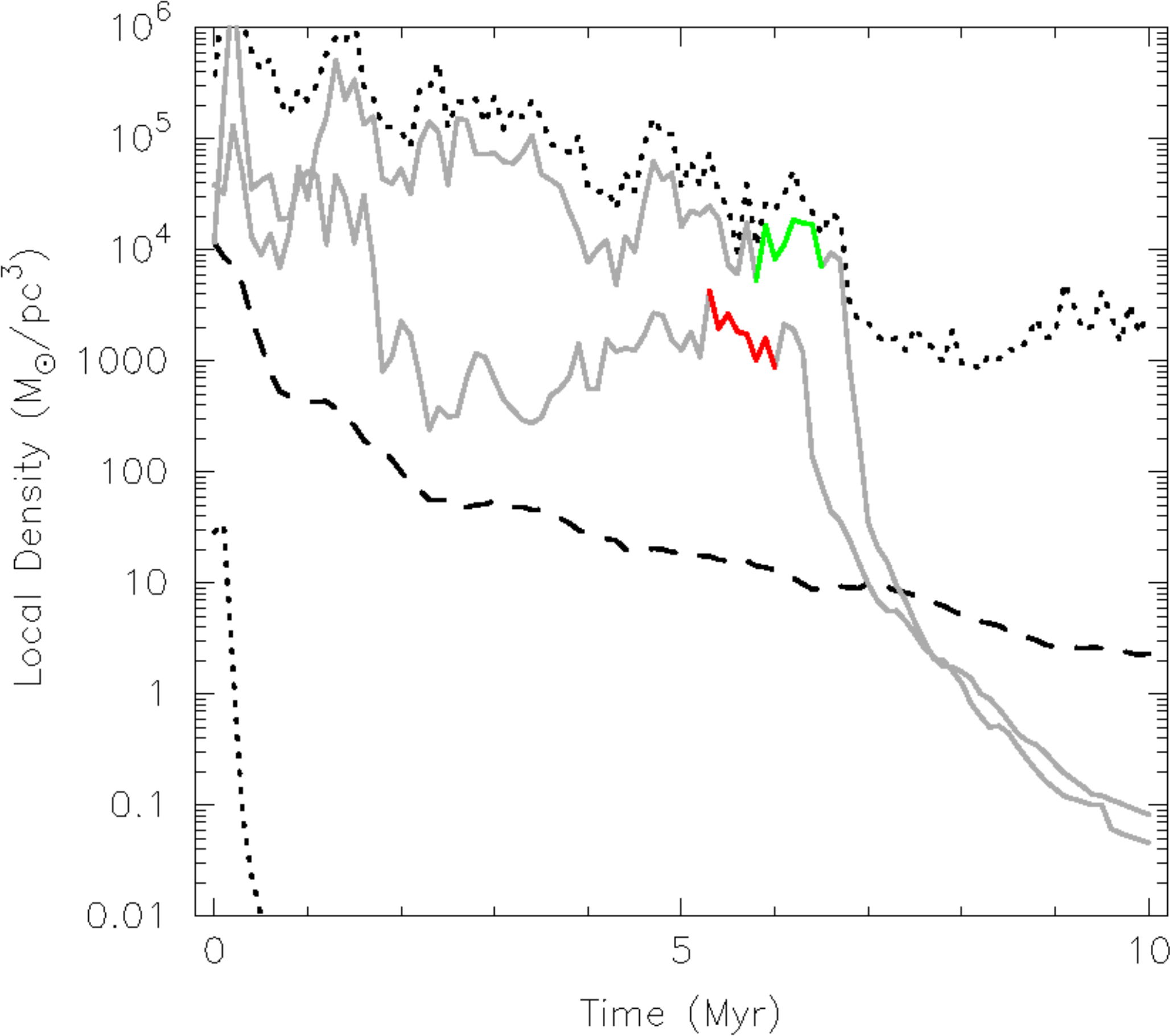}
\hspace*{-0.6cm}
\includegraphics[scale=0.38]{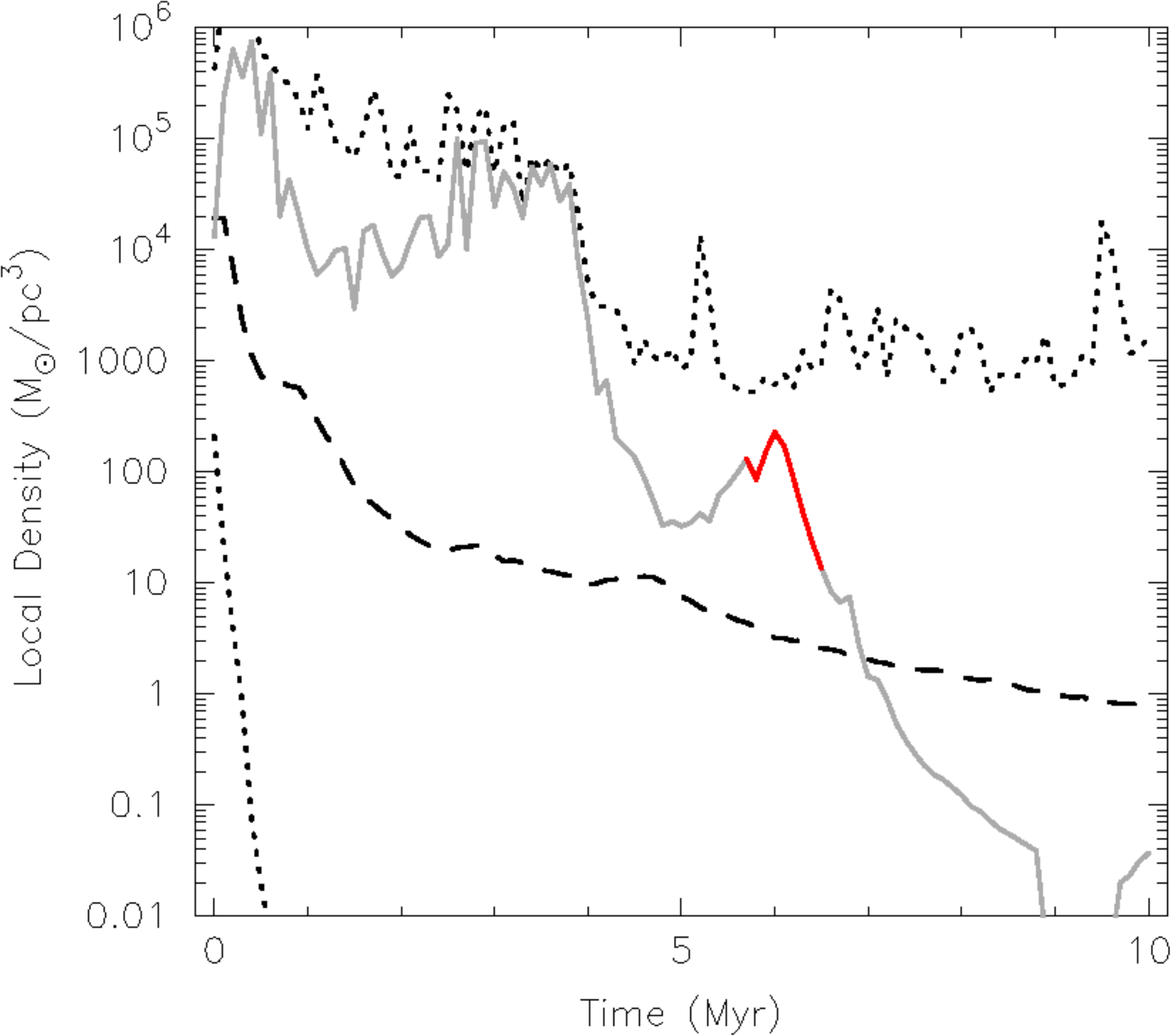}
\caption{Local densities around stars in simulated clusters. The dotted lines are the upper and lower density bounds of the cluster, whilst the dashed line is the median. The solid lines are the stars that will evolve into a WR star, with the coloured segments denoting the WR phase. In the upper panel, there are two WR stars in the cluster (red and green), both of which remain in dense regions during their lifetimes. However, in the lower panel, the WR star has moved into a moderately dense environment during its evolution (though the surroundings are still denser than the median of the cluster).}
\label{fig:cr}
  \end{center}
\end{figure}

\subsection{Dissolution of star clusters}

We now explore the possibility that WR stars appear to be isolated because their birth, or host star clusters have dissolved. 
Observations indicate that only 10\,per cent of star clusters survive beyond an age of 10\,Myr \citep{2003ARA&A..41...57L}. The 
exact reasons for this rapid destruction of star clusters is still debated. Clusters could be disrupted by the expulsion of gas 
leftover from the star formation process 
\citep{1978A&A....70...57T,1984ApJ...285..141L,1997MNRAS.284..785G,2007MNRAS.380.1589B,2018ApJ...863..171S}, although the 
effectiveness of this mechanism has been questioned \citep{2012MNRAS.419..841K}.

An alternative to gas expulsion is the rapid expansion of a star cluster through two-body and violent relaxation 
\citep{2014MNRAS.438..620P}, which has been shown to cause clusters to expand significantly, thereby also significantly reducing 
the stellar density \citep{2012MNRAS.425..450M,2012MNRAS.426L..11G,2012MNRAS.427..637P}. In this scenario, the rapid ($<$10\,Myr) 
dynamical expansion of clusters could result in stellar densities similar to the Galactic field 
\citep[$\sim$0.1\,M$_\odot$\,pc$^{-3}$,][]{2003AJ....126.2896K}, causing the WR star(s) to appear isolated.

To test this hypothesis, we perform $N$-body simulations of the evolution of star-forming regions with a range of initial 
conditions. We use the results to determine the median stellar density during the WR phase of the massive stars, and compare this 
to the local stellar density surrounding the WR stars. The simulations are modified versions of those presented in 
\citet{2014MNRAS.437..946P} and \citet{2014MNRAS.438..620P} and we refer the interested reader to those papers for a full 
description. However, we summarise the initial conditions here.

\begin{table}
  \centering
  \caption[bf]{The variation in initial conditions of our $N$-body simulations. We show the initial radius, fractal dimension, the resultant initial stellar density, the initial virial state and the figures showing the particular simulation. }
  \begin{tabular}{|p{.1\linewidth}|p{.15\linewidth}|p{.15\linewidth}|p{.1\linewidth}|p{.20\linewidth}}
    \hline
    Radius & Fractal dimension $D$ & Initial density (M$_\odot$\,pc$^{-3}$) & Virial state & Figure(s) \\
    \hline
    1\,pc & 1.6 & $10^4$ & Sub & Fig.~\ref{fig:cr} \\
    \hline
    1\,pc & 1.6 & $10^4$ & Super & Fig.~\ref{fig:expand} (upper panel) \\
    5\,pc & 1.6 & $10^2$ & Super & Fig.~\ref{fig:expand} (lower panel) \\
    \hline
    5\,pc & 2.0 & $10$ & Super & Fig.~\ref{fig:modexpand} \\
    \hline
  \end{tabular}
  \label{table:simulations}
\end{table}

\begin{figure}
  \begin{center}
\includegraphics[scale=0.38]{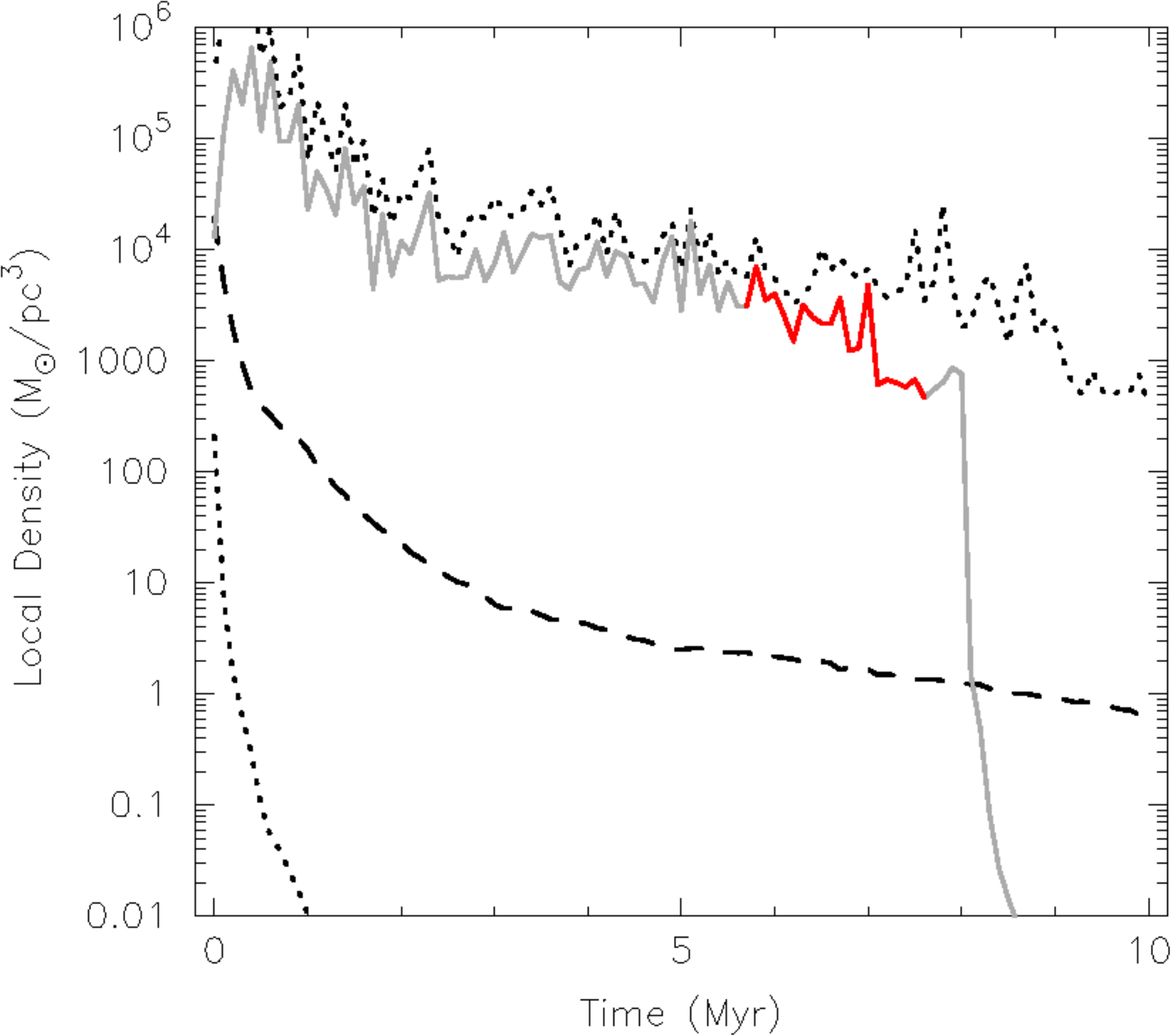}
\includegraphics[scale=0.38]{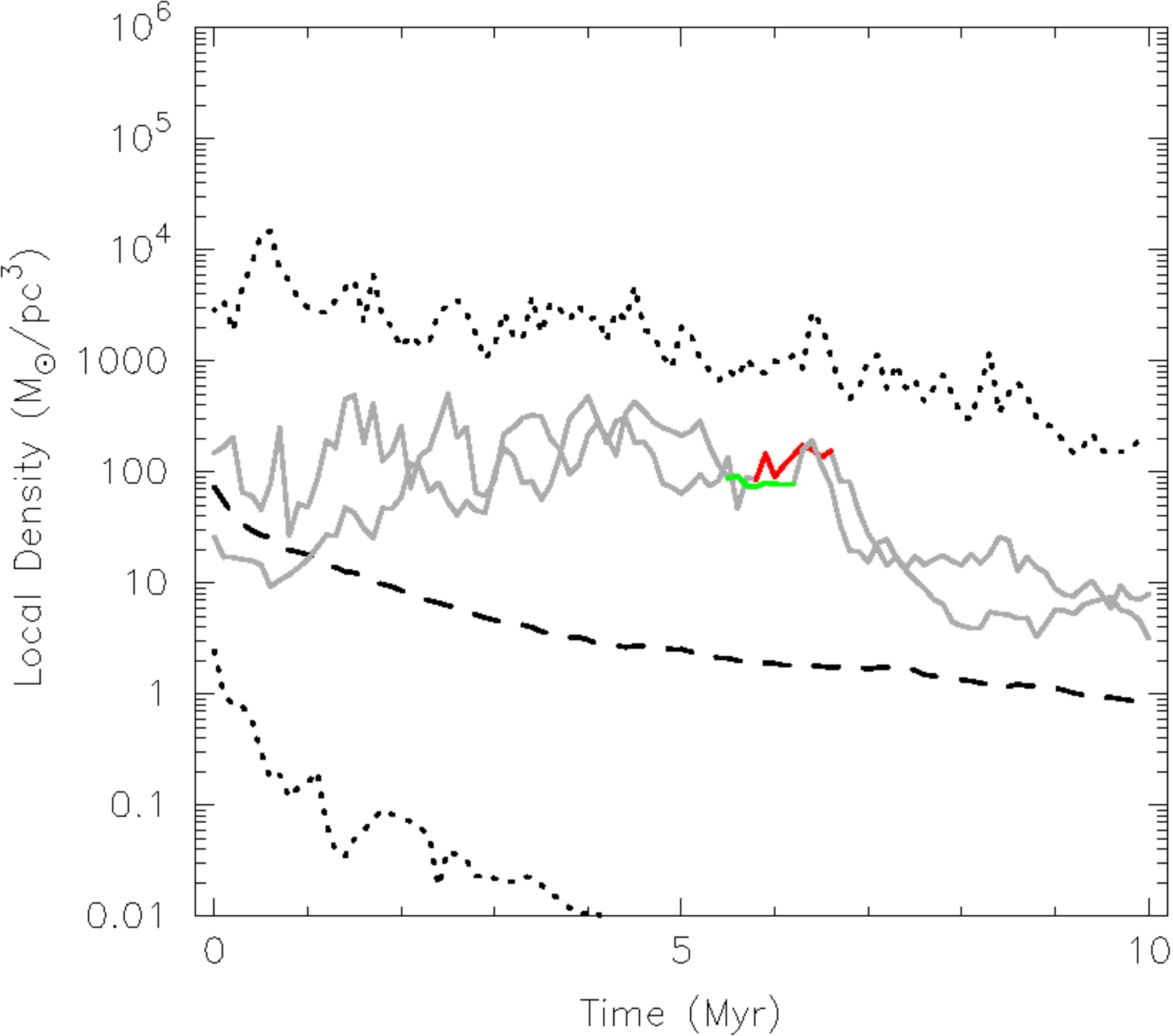}
\caption{Local densities around stars in expanding environments. The upper panel shows an initially dense, highly substructured simulation. The WR star remains in regions of high density due to mass segregation, before being ejected from the cluster. In the lower panel, a moderate density, but highly substructured simulation, the WR stars also remain in moderate or dense surroundings. These regions are still dense enough to be distinguished from the field.}
\label{fig:expand}
  \end{center}
\end{figure}

\begin{figure}
	\centering
  \includegraphics[width=\linewidth]{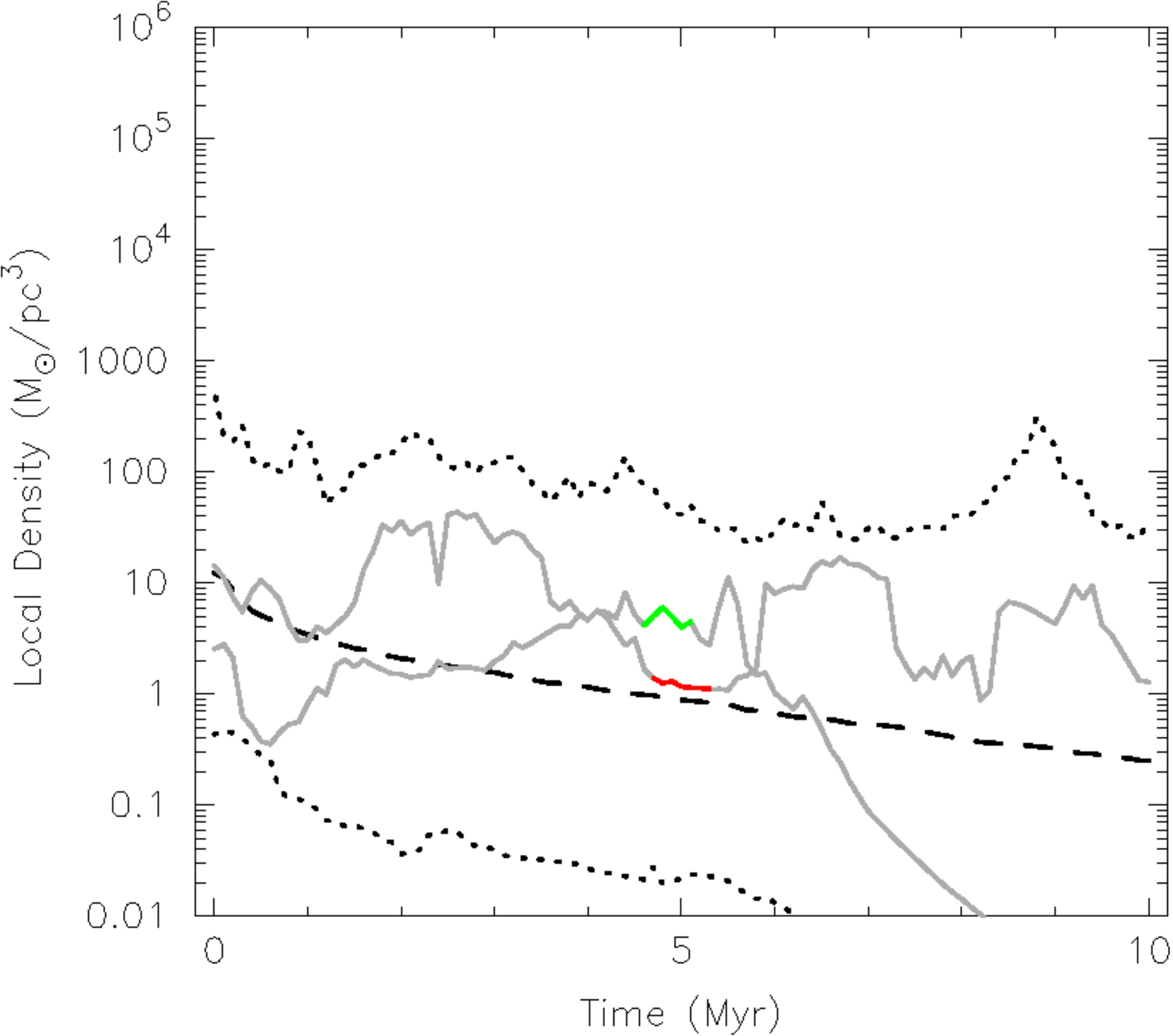}
  \caption{An expanding, low density, moderate substructure simulation. Here, the WR stars are in sparse environments and so appear to be isolated.} 
	\label{fig:modexpand}
\end{figure}

We draw 1500 stars randomly from the stellar IMF described in Eqn.~\ref{maschberger_imf}. Occasionally, this results in clusters 
with no sufficiently massive stars ($>$25\,M$_\odot$), but usually between one and five stars are massive enough to undergo a 
WR phase. We distribute these stars randomly in space and velocity within a fractal distribution \citep{2004A&A...413..929G}, which 
is the most straightforward way of creating the spatially and kinematically sub-structured initial conditions observed in young 
star-forming regions.

The amount of sub-structure in these fractals is set by the fractal dimension $D$; a star-forming region with high initial spatial 
and kinematic substructure has a low fractal dimension ($D = 1.6$), whereas a star-forming region with moderate amounts of 
substructure has a higher fractal dimension ($D = 2.0$). We run simulations with both degrees of substructure \citep[as 
observations also indicate a wide range of initial substructure in a star-forming region,][]{2004MNRAS.348..589C}. Simulations with 
a high amount of initial sub-structure have higher local stellar densities relative to those with less substructure. We also vary 
the \emph{global} density of the simulations by varying the initial radius of the region, which is either 1 or 5\,pc.

Finally, we vary the initial virial ratio. Our star-forming regions can be subvirial, which means they collapse into the potential 
well of the region and form a smooth, spherically symmetric star cluster. Following the formation of the star cluster, two-body 
relaxation dominates and the cluster expands. In other simulations, the star-forming regions are initially supervirial, which means 
they expand immediately.

A summary of the initial conditions of our $N$-body simulations, and the corresponding figure references, are given in 
Table~\ref{table:simulations}. In all simulations, we evolve the star-forming regions for 10\,Myr (i.e.\,\,long enough for the 
massive stars to undergo the WR phase before evolving into a stellar remnant). We include stellar evolution using the \texttt{SeBa} 
package \citep{1996A&A...309..179P} and the star-forming regions are evolved dynamically using \texttt{kira} 
\citep{1999A&A...348..117P}.

20 simulations were run for each combination of substructure, global density and initial virial ratio parameters. As we used the 
   same 1500 stars for each set of initial conditions, 13 out of 20 simulations contained WR stars regardless of the other initial 
   parameters.

Fig.~\ref{fig:cr} shows two of the initially subvirial simulations (i.e.\,\,they collapse to form a cluster). The cluster then 
expands via two-body relaxation. The median, upper and lower bounds of local density variation are plotted over time, alongside the 
density for stars with initial mass >25 \Msolar. The median local density falls during cluster dissolution, as expected. However, 
due to mass segregation \citep{2010MNRAS.407.1098A,2014MNRAS.438..620P}, 90\% of WR stars remain preferentially located in regions 
of high (>1000 \Msolar$\mathrm{pc^{-3}}$) density. These regions are still recognizable as clusters, whilst the outer regions have 
dissolved. This suggests that WR stars in isolated environments are unlikely to originate from dissolved clusters.

We can repeat this analysis for unbound and lower density star-forming regions, which form stellar associations. For initially 
expanding simulations, with high density and substructure, the WR stars tend to form in moderate ($\sim$100 
\Msolar$\mathrm{pc^{-3}}$ 45\% of WR stars) or high ($\sim$1000 \Msolar$\mathrm{pc^{-3}}$, 36\% of WR stars) density regions and 
remain in these density enhancements (e.g. upper panel of Figure ~\ref{fig:expand}).

Moderately dense but highly sub-structured simulations produced a similar result, with 77\% of WR stars remaining in moderately or 
highly dense environments (e.g. lower panel of Figure~\ref{fig:expand}). The remaining WR stars were located in low density environments ($\sim$10 \Msolar$\mathrm{pc^{-3}}$), which are comparable to the field. This implies that  such regions can produce WR stars that appear to be isolated, but that they are not the most common formation environment.

However, the expanding simulations with moderate substructure and initially low density in Figure ~\ref{fig:modexpand}, led to typical WR 
densities of around 1-10 \Msolar$\mathrm{pc^{-3}}$. To an observer, this is comparable to the field density and occurred because 
the moderate substructure and low-density prevented WR stars from becoming mass segregated. A corresponding collapsing version of 
this simulation produced WR stars in much more dense environments of 10-1000 \Msolar$\mathrm{pc^{-3}}$.

The final set of simulations suggest that WR stars primarily form in low density and sub-structured environments within 
associations. These regions would dissolve to field densities, via very gentle expansion, over the WR progenitor lifetime, which 
would make the WR star appear isolated.

Cygnus OB2 is an example of one such region. Based on the \citet{2014MNRAS.438..639W} surface density of 13.3 stars 
$\mathrm{pc^{-2}}$, the typical volume density of Cygnus OB2 (which is somewhat lower than the surface density) is $\sim$5 stars 
$\mathrm{pc^{-3}}$. With a typical IMF, this results in an average mass density which is similar to the field, at $\sim$1-10 
\Msolar$\mathrm{pc^{-3}}$.


\section{Discussion and conclusions} \label{sec:con}

\begin{figure}
	\centering
  \includegraphics[width=\linewidth]{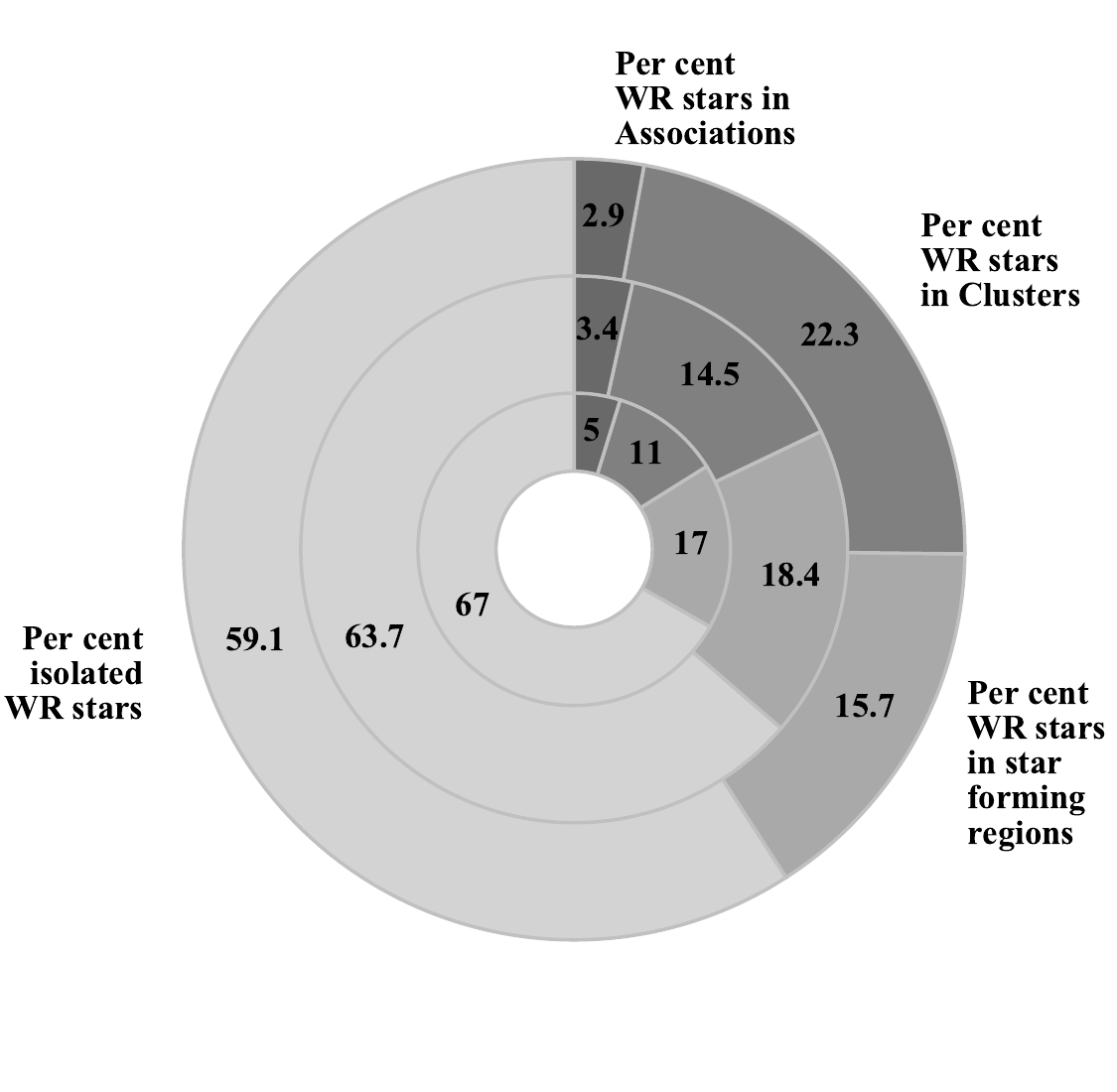}
  \caption{Doughnut chart showing the percentages of the WR stars 
with \textit{Gaia} DR2 distances in clusters, associations and star forming regions and isolated 
environments (inner, 379 stars), \textit{Gaia} plus embedded Galactic disk WR stars (middle, 553 stars), \textit{Gaia} plus all 
embedded WR stars (outer, 663 stars). } 
	\label{fig:pie_chart}
\end{figure}

We have exploited \textit{Gaia} DR2 proper motions and parallaxes to reassess the membership fraction of WR stars in clusters 
and OB associations within the Galactic disk. Only 16\% (61 of the 379) WR stars identified in \textit{Gaia} DR2 are 
confirmed members of clusters or OB associations, with a further 23 stars possible cluster/association members, plus 42 potential 
members of visibly obscured star-forming regions. The large distances and high visual extinctions of most WR stars precludes 
membership of known OB associations. Consequently, 67--84\% of the WR stars accessible to \textit{Gaia} DR2 are 
isolated,  in contrast to only 13\% of the Galactic O star population within a few kpc of the Sun. The fraction of
isolated WR stars within the Galactic disk is largely unchanged if WR stars inaccessible to \textit{Gaia} are considered (64--82\% 
of 553 WR stars). Once literature results for the WR populations within the Galactic Centre region are included too, 59--75\% of 663 Galactic WR stars are isolated. This is illustrated in Figure ~\ref{fig:pie_chart}.

Our results are broadly consistent with literature results for the membership of Galactic WR stars in clusters or associations 
\citep{1984A&AS...58..163L} but is much lower than their progenitor O stars, for which over 70\% of the 611 O stars in v3 of the 
Galactic O star Catalogue \citep{2013msao.confE.198M} are members of OB associations or star-forming regions. We explore the origin 
of the high field WR population by  undertaking simulations of star forming regions in which WR stars result from progenitors with 
$\geq$25 $M_{\odot}$.

We find that WR stars in low mass star forming regions lacking other massive stars contribute 8--15\% of isolated cases, of which WR11 (WC8+O) in the $\gamma$ Velorum group may be an example. Additionally, N-body simulations of clusters containing WR stars, 
reveal that as the median density falls, the outer regions of the cluster dissolve into the field. However, due to mass 
segregation, WR members remain in high density regions, which would appear as clusters. Only simulations of expanding, moderately sub-structured environments which are already low density, reproduced WR stars that appeared to be isolated. This suggests that most WR form 
in less dense associations, which are expanding from birth and dissolve to make the WR star appear isolated during its lifetime.

We conclude that only a subset of WR progenitors originate from dense, massive star clusters, such as NGC 3603 or Westerlund 
1, with a significant fraction from more modest open clusters, such as Collinder 232, NGC~6231 and Trumpler 16. Considering the global 
WR population of the Milky Way, 22\% are members of clusters, versus 40\% of Galactic O stars, indicating that up to half of 
massive stars are dynamically ejected from such clusters \citep{2016A&A...590A.107O, 2018MNRAS.480.2109D}.

From Paper I, we identify a minimum of 8\% for the runaway fraction of \textit{Gaia} DR2 WR stars owing  to being located more 
than 156 pc (three \hii\ region scale heights) from the Galactic mid-plane. This is in accord with a runaway fraction of  10--25\% for O 
stars according to \citet{1986ApJS...61..419G}. Although a significant fraction of massive stars are believed to be  dynamically 
ejected from star forming regions \citep{2016A&A...590A.107O} or disruption of binaries following a core-collapse  supernova 
\citep{2019A&A...624A..66R}, runaways are predicted to be extremely rare, in tension with the observed runaway rate.

Overall, based on the observed cluster and association membership fractions, and the simulations conducted, we propose
that the isolation of WR stars can be explained by the following scenario:

\begin{itemize}[leftmargin=*]  
  \item $\sim$20\% of WR stars form in rich open clusters, such as NGC 3603 (or other clusters in Carina), and remain in situ 
throughout their lives.
  \item $\sim$20\% of WR stars appear isolated because they have been ejected from their birth star-forming region, either through 
dynamical ejection or binary disruption.
    \item $\sim$10\% are isolated because they have formed in a low mass  ($\sim$100 \Msolar) region, containing only a single WR 
star and no OB stars. The remainder of the stellar population, aside from the WR star, is therefore too faint to be observed, 
which makes the WR star appear to be isolated (e.g. WR11 within the $\gamma$ Velorum group)
   \item $\sim$5\% of the WR population still reside in non-clustered OB associations/star-forming regions. These regions may be 
dissolving and have therefore not yet  reached field densities, or they may have started out slightly more dense than the typical 
WR star environment and are therefore taking longer to fully expand.
  \item The remaining 45\% of WR stars originate in low density, moderately sub-structured associations, which expand during 
the WR star lifetime to low densities ($\sim$1-10M$_\odot$\,pc$^{-3}$), which again makes the WR star appear isolated. An 
observational example of this environment is Cygnus OB2. 
\end{itemize}

To verify this scenario and better constrain the fraction of WR stars in each environment, future work should consider re-assessing 
the regions around known WR stars (in particular the stars we were not able to firmly classify in this study). This could be done 
using \textit{Gaia} astrometric data and clustering algorithms, to identify all possible members of the surrounding stellar 
population; therefore ensuring completeness that may not be present in literature membership lists (e.g \citealt{2019A&A...621A.115C}, see Section \ref{ssec:Gammavel}). The region could then be characterised to determine if it expanded from the moderately dense and sub-structured environments, that may comprise the majority of WR star formation sites.

In addition to establishing the environment of WR stars in the Galactic disk, we have also reassessed the distance to 
clusters/associations from \textit{Gaia} DR2, considering systematics (\citealt{gaia_pres}, \citeyear{2018A&A...616A...2L}) and random 
uncertainties. For those clusters host to WR stars, we used cluster distances, literature photometry and spectral types, to 
calculate the extinctions and then the luminosities of OB members, from which cluster ages were estimated using isochrones from 
\citet{2011A&A...530A.115B}.  Previous results are largely supported, in which young clusters ($\leq$ 2 Myr) host H-rich WN or 
Of/WN stars, intermediate age clusters (2--5 Myr) host classical WR stars, with older ($\geq$5 Myr) clusters, host to stars which could have been affected by binary evolution (e.g. rejuvenation following a stellar merger), but we do not see a large population of stars which may have evolved via the WR binary formation channel.

Finally, let us return to a topic mentioned in our introduction, namely whether ccSNe environments support WR stars as the primary progenitors of SE-SNe. Since the incidence of star-formation increases from the least stripped (II-P, II-L) to the most stripped (IIb, Ib, Ic) ccSNe \citep{2012ApJ...759..107K, 2012MNRAS.424.1372A, 2018A&A...613A..35K}, more massive progenitors are inferred for SE-SNe, especially for broad lined Ic SNe. By way of example, \citet{2013MNRAS.428.1927C} found 70$\pm$26\% of nearby type Ib/c ccSNe to be associated with a H\,{\sc ii} region, versus only 38$\pm$11\% for type II SNe at similar distances.

If WR stars are responsible for (some) SE-SNe, how can one reconcile the fact that $\geq$60\% of Galactic WR stars are not associated with star formation (Fig~\ref{fig:pie_chart}), yet the overwhelming majority of SE-SNe are associated with star forming regions? Recall that 
the typical distance to a Galactic WR star is 5 kpc (Paper I), versus a mean distance of 20 Mpc for stripped envelope SNe within large samples \citep{2012MNRAS.424.1372A}. 100 parsecs subtends over a degree at the distance of a typical Galactic WR star, versus 1 arcsec for SE-SNe. Consequently, for a Galactic WR star to be associated with a star-forming region, it needs to lie in the H\,{\sc ii} region, or in its close proximity (1--5 pc). In contrast, SE-SNe are flagged as being associated with a star-forming regions if they lie within 50--100 pc of the H\,{\sc ii} region.

The size of H\,{\sc ii} regions spans a wide range \citep{1984ApJ...287..116K, 2013MNRAS.428.1927C}, from  $\sim$1 pc for compact H\,{\sc ii} regions (e.g. M42/Orion Nebula Cluster) to $\sim$1 kpc for supergiant H\,{\sc ii} regions (e.g. NGC 5461 in M101). According to  fig.~8 of \citet{2018A&A...613A..35K}, 60\% of SE-SNe are associated with star forming regions whose H$\alpha$ luminosities are inferior to that of the Rosette Nebula (NGC~2264), with 30\% below M42. Physical dimensions of characteristic extragalactic H\,{\sc ii} regions are thus $\sim$10 pc, significantly smaller than the resolution of non-Adaptive Optics, ground-based observations.

Therefore, if one was to relax the condition that a Galactic WR star is associated with a star-forming region by an order of magnitude, from $\sim$5 parsec to $\sim$50 parsec, the statistics would of course be far greater than $\leq$40\%. However, such an exercise awaits the combination of more robust \textit{Gaia} parallaxes in future data releases and reliable distances to Galactic star-forming regions \citep{2019ApJ...885..131R}.

\section*{Acknowledgements}

GR wishes to thank the Science and Technology Facilities Council (STFC), for their financial support through the Doctoral Training Partnership. RJP acknowledges support from 
the Royal Society in the form of a Dorothy Hodgkin Fellowship.

We wish to thank the referee Dr Anthony Brown for his helpful comments and suggestions on the submitted manuscript. This work has made use of data from the European Space Agency (ESA) mission {\it Gaia} (\url{https://www.cosmos.esa.int/gaia}), processed by the {\it Gaia} Data Processing 
and Analysis Consortium (DPAC, \url{https://www.cosmos.esa.int/web/gaia/dpac/consortium}). Funding for the DPAC has been provided by national institutions, in particular the 
institutions participating in the {\it Gaia} Multilateral Agreement.

This publication also makes use of data products from the Two Micron All Sky Survey, which is a joint project of the University of Massachusetts and the Infrared Processing 
and Analysis Center/California Institute of Technology, funded by the National Aeronautics and Space Administration and the National Science Foundation.

This research has made use of the WEBDA database, operated at the Department of Theoretical Physics and Astrophysics of the Masaryk University and of the VizieR catalogue 
access tool, CDS, Strasbourg, France. The original description of the VizieR service was published in A\&AS 143, 23

This work would not be possible without the python packages Numpy (\citealt{numpy_book}, \citealt{doi:10.1109/MCSE.2011.37}), Pandas \citep{mckinney-proc-scipy-2010}, Scipy \citep{2020SciPy-NMeth} and Matplotlib \citep{2007CSE.....9...90H}.


\bibliographystyle{mnras}


\bibliography{gaia_class_pc2} 
\makeatletter
\relax
\def\mn@urlcharsother{\let\do\@makeother \do\$\do\&\do\#\do\^\do\_\do\%\do\~}
\def\mn@doi{\begingroup\mn@urlcharsother \@ifnextchar [ {\mn@doi@}
  {\mn@doi@[]}}
\def\mn@doi@[#1]#2{\def\@tempa{#1}\ifx\@tempa\@empty \href
  {http://dx.doi.org/#2} {doi:#2}\else \href {http://dx.doi.org/#2} {#1}\fi
  \endgroup}
\def\mn@eprint#1#2{\mn@eprint@#1:#2::\@nil}
\def\mn@eprint@arXiv#1{\href {http://arxiv.org/abs/#1} {{\tt arXiv:#1}}}
\def\mn@eprint@dblp#1{\href {http://dblp.uni-trier.de/rec/bibtex/#1.xml}
  {dblp:#1}}
\def\mn@eprint@#1:#2:#3:#4\@nil{\def\@tempa {#1}\def\@tempb {#2}\def\@tempc
  {#3}\ifx \@tempc \@empty \let \@tempc \@tempb \let \@tempb \@tempa \fi \ifx
  \@tempb \@empty \def\@tempb {arXiv}\fi \@ifundefined
  {mn@eprint@\@tempb}{\@tempb:\@tempc}{\expandafter \expandafter \csname
  mn@eprint@\@tempb\endcsname \expandafter{\@tempc}}}


\makeatother


\appendix


\bsp	
\label{lastpage}
\end{document}